\documentstyle[12pt]{article}
\advance\textheight by \topskip
\begin{document}
\def\lsim{\mathrel{\lower2.5pt\vbox{\lineskip=0pt\baselineskip=0pt
\hbox{$<$}\hbox{$\sim$}}}}
\def\gsim{\mathrel{\lower2.5pt\vbox{\lineskip=0pt\baselineskip=0pt
\hbox{$>$}\hbox{$\sim$}}}}
\def\gs{SU(2)_{\rm L} \times U(1)_{\rm Y}}
\def\lr{SU(2)_L \times SU(2)_R}
\def\lpr{SU(2)_{L+R}}
\def\sbs{{\cal SBS}}
\def\wh{{\cal W}}
\def\bh{{\cal B}}
\def\wtu{\wh^{\mu \nu}}
\def\wtd{\wh_{\mu \nu}}
\def\btu{\bh^{\mu \nu}}
\def\btd{\bh_{\mu \nu}}
\def\cl{{\cal L}}
\def\nll{\cl_{\rm NL}}
\def\ecl{\cl_{\rm EChL}}
\def\fpnl{\cl_{\rm FP}^{\rm NL}}
\def\fpl{\cl_{\rm FP}}
\def\msb{{\overline{\rm MS}}}
\def\mh{M_H}
\begin{titlepage}
\title{\Large{\bf Introduction to the Symmetry Breaking
Sector}\footnote{Lectures given at the XXIII International Meeting On
Fundamental Physics, Comillas, Santander, Spain, May 1995}}
\author{Mar\'{\i}a J.
Herrero\thanks{e--mail:herrero@delta.ft.uam.es; herrero@vm1.sdi.uam.es}
 \\[3mm] Departamento de F\'{\i}sica
Te\'orica\\ Universidad Aut\'onoma de Madrid\\ Cantoblanco,\
\ 28049 -- Madrid,\ \ Spain} \date{}
\maketitle
\def\baselinestretch{1.3}
\begin{abstract}
\noindent
{\normalsize
The basic ingredients of the Spontaneous Symmetry
Breaking Phenomenon and of the Higgs Mechanism are reviewed in these
lectures of pedagogical character. Some relevant topics related with the
breaking $\gs \rightarrow U(1)_{\rm em}$ are selected and discussed
here. A brief survey of the experimental Higgs particle searches and the
theoretical limits on $\mh$ are also presented. The main features of the
most popular models of symmetry breaking  beyond the Standard
Model are briefly considered. It includes a short summary of the Higgs
Sector in the Minimal SUSY Model, the basic ideas of Technicolor models
and a brief introduction to Strongly Interacting Scalar
sectors
and to the Effective Chiral Lagrangian Approach to the Electroweak
Theory.
 }
\end{abstract}

\vskip-17cm
\rightline{{\bf \large FTUAM Jan/96/1}} \vspace{1mm}

\rightline{{\bf \large hep-ph/9601--}} \vspace{1mm}

\rightline{{\bf \large January 1996}}

\end{titlepage}

\newpage

\section{Introduction}
One of the key ingredients of the Standard Model of electroweak
interactions (SM) \cite{gws} is the concept of Spontaneous Symmetry
Breaking (SSB) \cite{gold}, giving rise to Goldstone-excitations which
in turn can be
related to gauge boson mass terms. This procedure, ussually called Higgs
Mechanism \cite{higgs}, is necessary in order to describe the short
ranged weak interactions by a gauge theory without spoiling gauge
invariance. The discovery of the $W^{\pm}$ and $Z$ gauge bosons at CERN
in 1983 may be considered as the first experimental evidence of the
Spontaneous Symmetry Breaking Phenomenon in electroweak interactions
\cite{ssb,pich}.
In present and future experiments one  hopes to get insight into the
nature of this Symmetry Breaking Sector ($\sbs$) and this is one of the
main
motivations for constructing the next generation of accelerators. In
particular, it is the most exiciting challenge for the recently approved
LHC collider being built at CERN.

In the SM, the symmetry breaking is realized linearly by a scalar field
which acquires a non-zero vacuum expectation value. The resulting
physical spectrum contains not only the massive intermediate vector
bosons and fermionic matter fields but also the Higgs particle, a
neutral scalar field which has successfully escaped experimental
detection until now. The main advantage of the Standard Model picture of
symmetry breaking lies in the fact that an explicit and consistent
formulation exists, and any observable can be calculated perturbatively
in the Higgs self-coupling constant.  However, the fact that one can
compute in a model doesn't mean at all that this is the right one.

The concept of spontaneous electroweak symmetry breaking is more general
than the way it is usually implemented in the SM. Any alternative
$\sbs$ has a chance to replace the standard Higgs sector, provided it
meets the following basic requirements: 1) Electromagnetism
remains unbroken; 2) The full symmetry contains the electroweak gauge
symmetry; 3) The symmetry breaking occurs at about the energy scale
$v=(\sqrt{2}G_F)^{-\frac{1}{2}}=246\;GeV$ with $G_F$ being the Fermi
coupling constant.

In these lectures I will review all these basic ingredients of the
Symmetry Breaking Phenomenom in the Electroweak Theory, and I will
discuss some relevant topics related with this breaking. The lectures
aim to be of pedagogical character and they
are essentially addressed to young particle physicists without too
much theoretical background in Quantum Field Theory.
The lectures include a survey of experimental Higgs particle
searches and the present status of theoretical bounds on the Higgs mass.
Part of these lectures is devoted to present selected $\sbs$ beyond the
SM Higgs sector. It includes short introductions to: The Higgs sector
of the Minimal SUSY Model (MSSM), Technicolor models, Strongly
Interacting
Scalar sectors and the Electroweak Chiral Lagrangian (EChL).
These
Lectures do not pretend to treat exhaustively the subject of Electroweak
Symmetry Breaking nor to provide a complete set of references. I
appologize for possible (most probably) missing references.

\renewcommand\baselinestretch{1.3}
\section{The Phenomenon of Spontaneous Symmetry Breaking}
A simple definition of the phenomenon of spontaneous symmetry breaking
(SSB)\cite{ssb}
is as follows:

{\it A physical system has a symmetry that is spontaneously
broken if the interactions governing the dynamics of the system possess
such a symmetry but the ground state of this system does not}.

An illustrative example of this phenomenon is the infinitely extended
ferromagnet. For this purpouse, let us consider the system near the
Curie temperature $T_C$. It is described by an infinite set of
elementary spins whose interactions are rotationally invariant, but its
ground state presents two different situations depending on the value of
the temperature $T$.

{\bf Situation I}: $T>T_C$

\vspace{0.5cm}
The spins of the system are randomly oriented and as a consequence the
average magnetization vanishes: $\vec{M}_{\rm average}=0$. The ground
state with these disoriented spins is clearly rotationally invariant.

{\bf Situation II}: $T<T_C$

\vspace{0.5cm}
The spins of the system are all oriented parallely to some particular
but
arbitrary direction and the average magnetization gets a non-zero value:
$\vec{M}_{\rm average} \neq 0$ ({\it Spontaneous Magnetization}). Since
the
direction of the spins is arbitrary there are infinite possible ground
states, each one corresponding to one possible direction and all having
the same (minimal) energy. Futhermore, none of these states is
rotationally invariant since there is a privileged direction. This is,
therefore, a clear example of spontaneous symmetry breaking since the
interactions among the spins are rotationally invariant but the ground
state is not. More specifically, it is the fact that the system
'chooses' one among the infinite possible non-invariant ground states
what produces the phenomenon of spontaneous symmetry breaking.

On the theoretical side, and irrespectively of what could be the origen
of such a physical phenomenon at a more fundamental level, one can
parametrize this behaviour by means of a symple mathematical model. In
the case of the infinitely extended ferromagnet one of these models is
provided by the Theory of Ginzburg-Landau \cite{gl}. We present in the
following the basic ingredients of this model.

For $T$ near $T_C$, $\vec{M}$ is small and the free energy density
$u(\vec{M})$ can be approached by (here higher powers of $\vec{M}$ are
neglected):
\begin{eqnarray}
u(\vec{M})& = &(\partial_i \vec{M})(\partial_i \vec{M})+V(\vec{M})\;;\;
i=1,2,3 \nonumber\\
V(\vec{M})&=&\alpha_1(T-T_C)(\vec{M}.\vec{M})+\alpha_2(\vec{M}.\vec{M})^2
\;;\;\alpha_1,\alpha_2>0
\label{landau}
\end{eqnarray}
The magnetization of the ground state is obtained from the condition of
extremum:
\begin{equation}
\frac{\delta V(\vec{M})}{\delta M_i}=0\Rightarrow
\vec{M}.\left [ \alpha_1(T-T_C)+2\alpha_2(\vec{M}.\vec{M})\right ]=0
\label{min}
\end{equation}
There are two solutions for $\vec{M}$, depending on the value of $T$:

{\bf Solution I}:

\vspace{0.5cm}
If $T>T_C\; \Rightarrow \left [
\alpha_1(T-T_C)+2\alpha_2(\vec{M}.
\vec{M}) \right ]>0\;\Rightarrow \vec{M} =0$\\

The solution for $\vec{M}$ is the trivial one and corresponds to the
situation I described before where the ground state is rotational
invariant. The potential $V(\vec{M})$ has a symmetric shape with a
unique minimum at the origen $\vec{M}=0$ where $V(0)=0$. This is
represented in Fig.1 for the simplified bidimensional case,
$\vec{M}=(M_X,M_Y)$.\\
\vspace{8cm}
\begin{center}
{\bf Fig.1} The potential $V(\vec{M})$ in the symmetric phase
\end{center}
\vspace{0.5cm}
{\bf Solution II}:

\vspace{0.5cm}
If $T<T_C \; \Rightarrow \vec{M}=0$ is a local maximum and
eq.(\ref{min}) requires:
$$ \alpha_1(T-T_C)+2 \alpha_2(\vec{M}.\vec{M})=0 \Rightarrow
|\vec{M}|=\sqrt{\frac{\alpha_1(T_C-T)}{2\alpha_2}}$$
Namely, there are an infinite absolute minima having all the same
$|\vec{M}|$ above, but different direction of $\vec{M}$. This
corresponds to the situation II where the system has infinite possible
degenerate ground states which are not rotationally invariant. The
potential $V(\vec{M})$ has a 'mexican hat shape' as represented in
Fig.2 for the bidimensional case.

\newpage
${}^{}$
\vspace{8cm}
\begin{center}
{\bf Fig.2} The potential $V(\vec{M})$ in the spontaneously broken phase
\end{center}

\vspace{0.5cm}
Notice that it is the choice of the particular ground state what
produces,
for $T<T_C$, the spontaneous breaking of the rotational symmetry.

\section{Spontaneous Symmetry Breaking in Quantum Field Theory: QCD as
an example}
In the language of Quantum Field Theory, {\it a system is said to
possess
a symmetry that is spontaneously broken if the Lagrangian describing the
dynamics of the system is invariant under these symmetry
transformations, but the vacuum of the theory is not}. Here the vacuum
$|0>$ is the state where the Hamiltonian expectation value $<0|H|0>$ is
minimum.

For illustrative purposes we present in the following the particular
case of Quantum Chromodynamics (QCD) where  there is a symmetry, the
chiral symmetry, that is spontaneously broken \cite{donog}. For
simplicity
let us consider QCD with just two flavours. The Lagrangian is given by:
\begin{equation}
\cl_{\rm QCD}=-\frac{1}{2} TrG^{\mu\nu} G_{\mu\nu}+\sum_{u,d}
(i\bar{q}D_{\mu}\gamma^{\mu}q-m_q\bar{q}q)
\label{QCD}
\end{equation}
where,
\begin{eqnarray}
  G_{\mu\nu}&=&\partial_{\mu}A_{\nu}-\partial_{\nu}A_{\mu}-ig_s
  \left[A_{\mu},A_{\nu}\right]\nonumber \\
  D_{\mu}q &=& (\partial_{\mu}-ig_s A_{\mu})q \nonumber \\
 A_{\mu}&=&\sum_{a=1}^8 \frac{1}{2}A_{\mu}^a\lambda_a
\end{eqnarray}
It is easy to check that for $m_{u,d}=0$, $\cl_{QCD}$ has (apart from
the $SU(3)_C$ gauge symmetry) a global symmetry $\lr$, called chiral
symmetry,
that is defined by the following transformations:
\[ \Psi_L \rightarrow \Psi_L'=U_L \Psi_L \]
\[ \Psi_R \rightarrow \Psi_R'=U_R \Psi_R \]
where,
\[ \Psi = \left ( \begin{array}{c}u \\ d \end{array} \right ) \;;\;
\Psi_L=\frac{1}{2}(1-\gamma_5)\Psi\;;\;
\Psi_R=\frac{1}{2}(1+\gamma_5)\Psi \]
\[ U_L \in SU(2)_L \; ;\; U_R \in SU(2)_R \]

$U_L$ and $U_R$ can be written in terms of the 2x2 matrices $T_L^a$ and
$T_R^a$ ($a=1,2,3$) corresponding to the generators $Q_L^a$ and $Q_R^a$
of $SU(2)_L$ and $SU(2)_R$ respectively:
\[ U_L= \exp(-i\alpha_L^a T_L^a)\;;\;
   U_R= \exp(-i\alpha_R^a T_R^a)  \]

It turns out that the physical vacuum of QCD is not invariant under the
full chiral $\lr$ group but just under the subgroup $SU(2)_V= \lpr$ that
is
the well known isospin symmetry group. The transformations given by the
axial subgroup, $SU(2)_A$, do not leave the QCD vacuum invariant.
Therefore, QCD with $m_{u,d}=0$ has a chiral symmetry which is
spontaneously broken down to the isospin symmetry:
\[ \lr \rightarrow SU(2)_V \]

The fact that in nature $m_{u,d}\neq 0$ introduces an extra explicit
breaking of this chiral symmetry. Since the
fermion masses are small this explicit breaking is soft. The chiral
symmetry is not an exact but approximate symmetry of QCD.

One important question is still to be clarified. How do we know from
experiment that, in fact, the QCD vacuum is not $\lr$
symmetric?. Let us
assume for the moment that it is chiral invariant. We will see that this
assumption leads to a contradiction with experiment.

If $|0>$ is chiral invariant $\Rightarrow$
 \[  U_L|0>=|0>\;;\;U_R|0>=|0>
\Rightarrow T_L^a|0>=0;T_R^a|0>=0 \Rightarrow Q_L^a|0>=0\;;\;
Q_R^a|0>=0 \]

In addition, if $|\Psi>$ is an eigenstate of the
Hamiltonian and parity operator such that:
\[ H|\Psi>=E|\Psi>\;;\;P|\Psi>=|\Psi> \]
then,
\[
\exists
|\Psi'>= \frac{1}{\sqrt{2}}(Q_R^a-Q_L^a)|\Psi>\; / \; H|\Psi'>=E|\Psi'>
\;;\;P|\Psi'>=-|\Psi'> \]
In summary, if the QCD vacuum is chiral invariant there must exist pairs
of degenerate states in the spectrum,
the so-called parity doublets as $|\Psi>$ and $|\Psi'>$,
which are related by a chiral transformation
and have opposite parities. The absence of such parity doublets in the
hadronic spectrum indicates that the chiral symmetry must be
spontaneously broken. Namely, there must exist some generators $Q^a$ of
the
chiral group such that $Q^a|0>\neq 0$. More specifically, it can be
shown that these generators are the three $Q^a_5 \; (a=1,2,3)$ of the
axial
group, $SU(2)_A$. In conclusion, the chiral symmetry breaking pattern in
QCD is $\lr \rightarrow SU(2)_V$ as announced.

\section{Goldstone Theorem}
One of the physical implications of the spontaneous symmetry breaking
phenomenom is the appearance of massless modes. For instance, in the
case of the infinitely extended ferromagnet and below the Curie
temperature there appear modes connecting the different possible ground
states, the so-called spin waves.

The general situation in Quantum Fied Theory is described by the
Goldstone Theorem \cite{gold}:

{\it If a Theory has a global  symmetry of the Lagrangian which is not a
symmetry of the vacuum then there must exist one massless
boson, scalar or pseudoscalar, associated to each
generator which does not annihilate the vacuum and having its same
quantum numbers. These modes are referred to as Nambu-Goldstone bosons
or simply as Goldstone bosons.}

Let us return to the example of QCD. The breaking of the chiral symmetry
is characterized by $Q^a_5|0>\neq 0\; (a=1,2,3)$. Therefore, according
to Goldstone Theorem, there must exist three massless Goldstone
bosons, $\pi^a(x)\;\; a=1,2,3$, which are pseudoscalars. These bosons
are identified with the three physical pions.

The fact that pions have $m_{\pi}\neq 0$ is a consequence of the soft
explicit breaking in $\cl_{QCD}$ given by $m_q \neq 0$. The fact that
$m_{\pi}$ is small and that there is a large gap between this mass and
the rest of the hadron masses can be seen as another manifestation of
the spontaneous chiral symmetry breaking with the pions being the
pseudo-Goldstone bosons of this breaking.

\section{Dynamical Symmetry Breaking}
In the previous sections we have seen the equivalence between the
condition $Q^a |0> \neq 0$ and the non-invariance of the vacuum under
the symmetry transformations generated by the $Q^a$ generators:
\[ U|0> \neq |0> \;;\; U= \exp(i\epsilon^a Q^a) \]

In Quantum Field Theory, it can be shown that an alternative way of
characterizing the phenomenom of SSB is by
certain field operators that have non-vanishing vacuum
expectation values (v.e.v.).
\[ {\rm SSB} \Longleftrightarrow \exists \Phi_j /
<0|\Phi_j|0> \neq 0 \]

This non-vanishing v.e.v. plays the role of the order parameter
signaling the existence of a phase where the symmetry of the vacuum is
broken.

There are several possibilities for the nature of this field operator.
In particular, when it is a composite operator which represents a
composite state being produced from a strong underlying dynamics, the
corresponding SSB is said to be a dynamical symmetry breaking. The
chiral symmetry breaking in QCD is one example of this type of breaking.
The non-vanishing chiral condensate made up of a quark and an anti-quark
is the order paremeter in this case:
\[ <0|\bar{q}q|0> \neq 0 \Rightarrow \lr \rightarrow SU(2)_V \]
The strong interactions of $SU(3)_C$ are the responsible for creating
these $\bar{q}q$ pairs from the vacuum and, therefore, the value  of
the condensate $<0|\bar{q}q|0>$ should, in principle, be calculable from
QCD.

It is interesting to mention that this type of symmetry
breaking can happen similarly in more general $SU(N)$ gauge theories.
The corresponding gauge couplings become sufficiently strong at large
distances
and allow for spontaneous breaking of their additional chiral-like
symmetries. The corresponding order paremeter is also a chiral
condensate: $<0|\overline{\Psi}\Psi|0>\neq 0$.
\section{The Higgs Mechanism}
The Goldstone Theorem is for theories with spontaneously broken global
symmetries but does not hold for gauge theories. When a spontaneous
symmetry breaking takes place in a gauge theory the so-called Higgs
Mechanism operates \cite{higgs}:

{\it The would-be Goldstone bosons associated to the global symmetry
breaking do not manifest
explicitely in the physical spectrum but instead they 'combine' with the
massless gauge bosons and as result, once the spectrum of the theory is
built up on the asymmetrical vacuum, there appear massive vector
particles. The number of vector bosons that acquire a mass is
precisely equal to the number of these would-be-Goldstone bosons}.

There are three important properties of the Higgs Mechanism for
'mass generation' that are worth mentioning:
\begin{itemize}
\item[1.-]
It respects the gauge symmetry of the Lagrangian.
\item[2.-]
It preserves the total number of polarization degrees.
\item[3.-]
It does not spoil the good high energy properties nor the
renormalizability of the massless gauge theories \cite{hooft}.
\end{itemize}

We now turn to the case of the Standard Model (SM) of Electroweak
Interactions \cite{gws,pich}. We will see in the following how the Higgs
Mechanism
is implemented in the $\gs$ Gauge Theory in order to generate a mass for
the  weak gauge bosons, $W^{\pm}$ and $Z$.

The following facts must be
considered:
\begin{itemize}
\item[1.-]
The Lagrangian of the SM is gauge $\gs$ symmetric.
Therefore, anything we wish to add must preserve this symmetry.
\item[2.-]
We wish to generate masses for the three gauge bosons $W^{\pm}$ and $Z$
but not for the photon, $\gamma$. Therefore, we need three
would-be-Goldstone bosons, $\phi^+$, $\phi^-$ and $\chi$, which will
combine with the three massless gauge bosons of the $\gs$ symmetry.
\item[3.-]
Since $U(1)_{\rm em}$ is a symmetry of the physical spectrum, it must be
a symmetry of the vacuum of the Electroweak Theory.
\end{itemize}

From the above considerations we conclude that in order to implement
the Higgs Mechanism in the Electroweak Theory we need to introduce
'ad hoc' an
additional system  that interacts with the gauge sector
in a $\gs$ gauge invariant manner and whose self-interactions, being
also introduced 'ad hoc', must produce the wanted breaking, $\gs
\rightarrow U(1)_{\rm em}$, with the three associated would-be-Goldstone
bosons $\phi^+$, $\phi^-$ and $\chi$.
This sytem is the so-called Symmetry Breaking Sector ($\sbs$).

\section{The Symmetry Breaking Sector of the Electroweak Theory}
In this section we introduce and justify the simplest choice for the
$\sbs$ of the Electroweak Theory.

Let $\Phi$ be the additional system providing the $\gs \rightarrow
U(1)_{\rm em}$ breaking. $\Phi$ must fulfil the following conditions:
\begin{itemize}
\item[1.-]
It must be a scalar field so that the above breaking preserves Lorentz
invariance.
\item[2.-]
It must be a complex field so that the Hamiltonian is hermitian.
\item[3.-]
It must have non-vanishing weak isospin and hypercharge in order to
break $SU(2)_{\rm L}$ and $U(1)_{\rm Y}$. The assignment of quantum
numbers and
the choice of representation of $\Phi$ can be done in many ways. Some
possibilities are:
\subitem -  Choice of a non-linear representation: $\Phi$
transforms non-linearly under $\gs$.
\subitem -  Choice of a linear representation: $\Phi$ transforms
linearly under $\gs$. The simplest linear representation is a complex
doublet. Alternative choices are: complex triplets, more than one
doublet, etc. In particular, one may choose two complex doublets $H_1$
and $H_2$ as in the Minimal Supersymmetric Standard Model.
\item[4.-]
Only the neutral components of $\Phi$ are allowed to acquire a
non-vanishing v.e.v. in order to preserve the $U(1)_{\rm em}$ symmetry
of the vacuum.
\item[5.-]
The interactions of $\Phi$ with the gauge and fermionic sectors
must be introduced in a gauge invariant way.
\item[6.-]
The self-interactions of $\Phi$ given by the potential $V(\Phi)$ must
produce the wanted breaking which is characterized in this case by
$<0|\Phi|0> \neq 0$. $\Phi$ can be, in principle, a fundamental or a
composite field.
\item[7.-]
If we want to be predictive from low energies to very high energies the
interactions in $V(\Phi)$ must be renormalizable. If instead one
renounces
to the predictivity at such high energies there is an alternative
possibility: The SM could be considered as an effective theory of some
other fundamental theory which opeartes at much higher energies. In that
case the emerging effective potential $V_{\rm eff}(\Phi)$ could be
non-renormalizable
and its 'predictivity' must be restricted just to low energies.
\end{itemize}
By taking into account the above seven points one is led to the
following simplest choice for the system $\Phi$ and the Lagrangian of
the $\sbs$ of the Electroweak Theory:
\begin{eqnarray}
\cl_{\rm SBS}& =& (D_\mu \Phi)^\dagger (D^\mu \Phi) -V(\Phi) \nonumber
\\
V(\Phi)&=&- \mu^2\Phi^\dagger \Phi + \lambda (\Phi^\dagger \Phi)^2
\;;\;\lambda>0
\label{lsbs}
\end{eqnarray}
where,
\begin{eqnarray}
\Phi & = & \left(
\begin{array}{c}\phi^+ \\
 \phi_0 \end{array}\right)
\nonumber\\[2mm]
D_\mu \Phi &  = &  ( \partial_\mu - \frac{1}{2} i g
\vec{\tau}\cdot\vec{W}_\mu
-\frac{1}{2} i g' B_\mu) \Phi . \label{SMS}
\end{eqnarray}
Here $\Phi$ is a fundamental complex doublet with hypercharge
$Y(\Phi)=1$ and $V(\Phi)$ is the simplest renormalizable potential.
$\vec{W}_{\mu}$ and $B_{\mu}$ are the gauge fields of $SU(2)_{\rm L}$
and
$U(1)_{\rm Y}$ respectively and $g$ and $g'$ are the corresponding gauge
couplings.

It is interesting to notice the similarities with the
Ginzburg-Landau Theory. Depending on the sign of the mass parameter
($-\mu^2$), there are two possibilities for the v.e.v. $<0|\Phi|0>$ that
minimizes the potential $V(\Phi)$,
\begin{itemize}
\item[1)]
$(-\mu^2)>0$: The minimum is at:
\[ <0|\Phi|0>=0 \]
The vacuum is $\gs$ symmetric and therefore no symmetry breaking occurs.
\item[2)]
$(-\mu^2)<0$: The minimum is at:
\[|<0|\Phi|0>|=\left(
\begin{array}{c} 0 \\
 \frac{v}{\sqrt{2}} \end{array}\right)\;\;;\;\; {\rm arbitrary}\;\;
  arg\;\Phi \;\;;\;\;
v\equiv \sqrt{\frac{\mu^2}{\lambda}} \]

Therefore, there are infinite degenerate vacua corresponding to infinite
posssible values of $arg\;\Phi$. Either of these vacua is $\gs$
non-symmetric and $U(1)_{\rm em}$ symmetric. The breaking $\gs
\rightarrow
U(1)_{\rm em}$ occurs once a particular vacuum is chosen. As usual, the
simplest choice is taken:
\[<0|\Phi|0> \equiv \left(
\begin{array}{c} 0 \\
 \frac{v}{\sqrt{2}} \end{array}\right)\;;\; arg\; \Phi \equiv 0\;;\;
v \equiv \sqrt{\frac{\mu^2}{\lambda}}  \]
\end{itemize}
The two above symmetric and non-symmetric phases of
the Electroweak Theory are clearly similar to the two phases of the
ferromagnet that we have described within the Ginzburg Landau
Theory context. In the SM, the field $\Phi$ replaces the magnetization
$\vec{M}$ and the potential $V(\Phi)$ replaces $V(\vec{M})$. The SM
order papameter is, consequently, $<0|\Phi|0>$. In the symmetric phase
$V(\Phi)$ is as in Fig.1, whereas in the non-symmetric phase it is as in
Fig.2.

Another interesting aspect of the Higgs Mechanism, as we have already
mentioned, is that it preserves the total number of polarization
degrees. Let us make the counting in detail:
\begin{itemize}
\item[1)] {\bf Before SSB} \\
4 massless gauge bosons: $W^{\mu}_{1,2,3}, B^{\mu}$ \\
4 massless scalars: The 4 real components of $\Phi$,
$(\phi_1,\phi_2,\phi_3,\phi_4)$ \\

Total number of polarization degrees $= 4\times 2 + 4 = 12$
\item[2)] {\bf After SSB} \\
3 massive gauge bosons: $W^{\pm},Z$ \\
1 massless gauge boson: $\gamma$ \\
1 massive scalar: $H$ \\

Total number of polarization degrees: $3\times 3+1\times 2+1=12$
\end{itemize}
Furthermore, it is important to realize that one more degree than needed
is introduced into the theory from the beginning. Three of the real
components of $\Phi$, or similarly
$\phi^{\pm} \equiv \frac{1}{\sqrt 2}(\phi_1 \mp i \phi_2)$ and
$\chi=\phi_3$,
are the needed would-be Goldstone bosons and the fourth one $\phi_4$ is
introduced just to complete the complex doublet. After the symmetry
breaking, this extra degree translates into the apparition in the
spectrum of an extra massive scalar particle ,
the Higgs boson particle $H$.

Before ending this section, we would like to address the following
important question: Is it possible the Higgs Mechanism without the Higgs
particle?.
From the previous counting we see that, strictly speaking, it is not
needed to implement the symmetry breaking. The Higgs Mechanism does
require just the three would-be Goldstone bosons in order to generate
the masses of the $W^{\pm}$ and $Z$ gauge bosons. Two more questions
then
arise. How must be introduced these three scalars? and, what are the
consequences of eliminating the Higgs particle?. It turns out that the
only possiblity of introducing the minimal number of scalars, that is
three, is by means of a non-linear representation. One example is:
\[U  \equiv  \displaystyle{\exp\left( {i \;
\frac{\vec{\tau}\cdot\vec{\phi}}{v}}\right)},\;\;\;
v  = 246 \;{\rm GeV}, \;\;\; \vec{\phi} = (\phi^1,\phi^2,\phi^3) \]
$U$ is a $2 \times 2$ unitary matrix and  transforms linearly under
$\gs$:
\[ U \rightarrow g_L U g_R^+\;;\; g_L \in SU(2)_L\;,\;g_R \in U(1)_Y\]
However, the would-be Goldstone boson fields transform non-linearly:
\[ \vec{\Phi}\rightarrow F(\vec{\Phi}) \]
with $F$ a non-linear function.

Moreover, one may build a potential in terms of the field $U$ and its
derivatives, $V(U,\partial_{\mu}U)$, such that it produces the wanted
breaking $\gs \rightarrow U(1)_{\rm em}$. It can be shown that the model
in which one replaces the standard $V(\Phi)$ by this
$V(U,\partial_{\mu}U)$
fulfils the
requirements 1 to 6 presented before. Of course,
The derivatives in the later should be replaced by the corresponding
covariant derivatives in order to get gauge invariance of the full
Lagrangian.

The drawback of this model is that it fails in condition 7. In
contrast to the standard $V(\Phi)$, $V(U,\partial_{\mu}U)$ is
non-renormalizable and therefore it cannot be predictive to very high
energies. In conclusion, if we want to implement the Higgs Mechanism
without the Higgs particle we must renounce to the renormalizability of
the $\sbs$ and we need to build a sensible low energy effective
theory and define a way to deal with non-renormalizable interactions.

\section{The particle spectra of the Electroweak Theory}
 In order to get the particle spectra and the particle masses we first
write down the full SM Lagrangian which is $\gs$ gauge
invariant:
\begin{equation}
\cl_{\rm SM}=
\cl_{\rm YM}+\cl_{\Psi}+\cl_{\rm SBS}+\cl_{\rm YW}
\label{lsm}
\end{equation}
Here, $\cl_{\rm YM}$, $\cl_{\Psi}$, $\cl_{\rm SBS}$ and $\cl_{\rm YW}$
are the Lagrangians of the Yang Mills fields, the fermionic fields, the
$\sbs$ and the Yukawa interactions respectively:
\begin{eqnarray}
\cl_{\rm YM}&=&\frac{1}{2} Tr \left( \wtd \wtu + \btd \btu \right) +
\cl_{\rm GF} + \cl_{\rm FP} \\[2mm]
\cl_{\Psi}&=&\sum_{\Psi}i \overline{\Psi}\gamma^{\mu}D_{\mu}\Psi \\[2mm]
\cl_{\rm SBS}& =& (D_\mu \Phi)^\dagger (D^\mu \Phi)
+ \mu^2\Phi^\dagger \Phi - \lambda (\Phi^\dagger \Phi)^2 \\[2mm]
\cl_{\rm YW}&=& \lambda_e\bar{l}_L\Phi e_R+
\lambda_u\bar{q}\widetilde{\Phi}u_R+\lambda_d\bar{q}_L\Phi d_R+h.c.+
2^{nd}\; {\rm and}\; 3^{rd}\;{\rm families}
\end{eqnarray}
where,
$\cl_{\rm GF}$ and $\cl_{\rm FP}$ are the gauge fixing and Faddeev-Popov
terms respectively that we omit here for brevity.

The field strength tensors are,
\begin{eqnarray}
\wtd & \equiv &  \partial_\mu \wh_\nu - \partial_\nu \wh_\mu -
g [\wh_\mu, \wh_\nu],\nonumber\\[2mm]
\btd & \equiv &  \partial_\mu \bh_\nu - \partial_\nu \bh_\mu
\label{tensors}
\end{eqnarray}
and the fields are given by,
\begin{eqnarray}
 \wh_\mu & \equiv & \frac{ -i}{2}\;
\vec{W}_\mu \cdot \vec{\tau}\;;\;\;
\bh_\mu  \equiv  \frac{ -i}{2} \; B_\mu \;
\tau^3 \nonumber \\
l_L& = & \left ( \begin{array}{c}\nu_L \\ e_L \end{array}\right )\;;\;\;
q_L =  \left (\begin{array}{c}u_L \\ d_L \end{array}\right ) \nonumber
\\
\Phi & = & \left(
\begin{array}{c}\phi^+ \\
 \phi_0 \end{array}\right)\;;\;\;
\widetilde{\Phi}=i\tau_2\Phi^*=\left(\begin{array}{c}
\phi^*_0\\-\phi^-\end{array}\right)
\label{fields}
\end{eqnarray}
The following steps summarize the procedure to get the spectrum from
$\cl_{\rm SM}$:
\begin{itemize}
\item[1.-] A non-symmetric vacuum must be fixed. Let us choose, for
instance,
  \[<0|\Phi|0>=\left( \begin{array}{c} 0\\ \frac{v}{\sqrt{2}}
  \end{array}\right ) \]
\item[2.-] The physical spectrum is built by performing 'small
oscillations' around this vacuum. These are parametrized by,
\[\Phi(x)=\exp{\left ( i\frac{\vec{\xi}(x)\vec{\tau}}{v}\right ) }
\left(\begin{array}{c}
0\\ \frac{v+H(x)}{\sqrt{2}}\end{array}\right ) \]
where $\vec{\xi}(x)$ and $H(x)$ are 'small' fields.
\item[3.-] In order to eliminate the unphysical fields we make the
following gauge transformations:
\begin{eqnarray}
\Phi'&=&U(\xi)\Phi=\left(\begin{array}{c}0\\ \frac{v+H}
{\sqrt{2}}\end{array}\right)\;;\;
U(\xi)=\exp\left(-i
\frac{\vec{\xi}\vec{\tau}}{v}\right) \nonumber\\
l_L'&=&U(\xi)l_L\;;\;e_R'=e_R\;;\;q_L'=U(\xi)q_L
\;;\;u_R'=u_R\;;\;d_R'=d_R \nonumber\\
\left(\frac{\vec{\tau}\cdot\vec{W}'_{\mu}}{2}\right)&=&
U(\xi)\left(\frac{\vec{\tau}\cdot\vec{W}_{\mu}}{2}\right) U^{-1}(\xi)
-\frac{i}{g}(\partial_{\mu}U(\xi))U^{-1}(\xi)\;;\;B'_{\mu}=B_{\mu}
\end{eqnarray}
\item[4.-] Finally, the weak eigenstates are rotated to the mass
eigenstates which define the physical gauge boson fields:
\begin{eqnarray}
W^\pm_\mu & = & \frac{W'^1_\mu \mp i W'^2_\mu}{\sqrt 2},
\nonumber\\[2mm] Z_\mu & = & c\; W'^3_\mu - s\; B'_\mu
\nonumber, \\[2mm] A_\mu & = & s\; W'^3_\mu + c\; B'_\mu,
\end{eqnarray}
where, $c\equiv \cos \theta_W$, $s\equiv \sin \theta_W$ and $\theta_W$
is the weak angle defined by $\tan\theta_W\equiv\frac{g'}{g}$.
\end{itemize}
It is now straightforward to read the masses from the following
terms of $\cl_{\rm SM}$:
\begin{eqnarray}
(D_{\mu}\Phi')^{\dagger}(D^{\mu}\Phi')&=&\left(\frac{g^2v^2}{4}\right)
W^+_{\mu}W^{\mu -}+\frac{1}{2}\left(\frac{(g^2+g'^2)v^2}{4}\right)
Z_{\mu}Z^{\mu}+...\nonumber \\
V(\Phi')&=&\frac{1}{2}(2\mu^2)H^2+...\nonumber \\
\cl_{\rm YW}&=&
\left(\lambda_e \frac{v}{\sqrt{2}}\right)\bar{e}'_R e'_R+
\left(\lambda_u \frac{v}{\sqrt{2}}\right)\bar{u}'_R u'_R+
\left(\lambda_d \frac{v}{\sqrt{2}}\right)\bar{d}'_R d'_R+...
\end{eqnarray}
and get finally the tree level predictions:
\begin{eqnarray}
M_W&=&\frac{gv}{2}\;;\;M_Z=\frac{\sqrt{g^2+g'^2}v}{2}\nonumber \\
M_H&=&\sqrt{2}\mu\nonumber\\
m_e&=&\lambda_e\frac{v}{\sqrt{2}}\;;\;
m_u=\lambda_u\frac{v}{\sqrt{2}}\;;\;
m_d=\lambda_d\frac{v}{\sqrt{2}}\;;...
\label{tree}
\end{eqnarray}
where,
\[ v=\sqrt{\frac{\mu^2}{\lambda}}\;;\;g=\frac{e}{s}\;;\;
g'=\frac{e}{c}\]
Some comments are in order.
\begin{itemize}
\item[-] All masses are given in terms of a unique mass parameter $v$
and the couplings $g$, $g'$, $\lambda$, $\lambda_e$, etc..
\item[-] The interactions of $H$ with fermions and with gauge bosons are
proportional to the gauge couplings and to the corresponding particle
masses:
\[ f\bar{f}H\;:\;-i\frac{g}{2}\frac{m_f}{M_W}\;;\;\;\;
W^+_{\mu}W^-_{\nu}H\;:\;igM_Wg_{\mu\nu}\;;\;\;\;
Z_{\mu}Z_{\nu}H\;:\;\frac{ig}{c}M_Zg_{\mu\nu}\]
\item[-] The v.e.v. $v$ is determined experimentally form $\mu$-decay.
By identifying the predictions of the partial width $\Gamma(\mu
\rightarrow
\nu_{\mu}
\bar{\nu}_e e)$ in the SM to low energies ($q^2<<M_W^2$) and in the V-A
Theory one gets,
\[\frac{G_F}{\sqrt{2}}=\frac{g^2}{8 M_W^2}=\frac{1}{2 v^2} \]
where,
\[ G_F= (1.16639\pm 0.00002)\times 10^{-5}\;GeV^{-2} \]
And from here,
\[v=(\sqrt{2} G_F)^{-\frac{1}{2}}= 246\;GeV\]
\item[-] The values of $M_W$ and $M_Z$ were anticipated successfully
quite before they were measured in experiment. The input
parameters were $\theta_W$, the fine structure constant $\alpha$ and
$G_F$. Before LEP these were the best measured electroweak parameters.
\item[-] In contrast to the gauge boson sector, the Higgs boson mass
$M_H$ and the Higgs self-coupling $\lambda$ are completely undetermined
in the SM.
\item[-] The hierarchy in the fermion masses is also
completely undetermined in the SM.
\end{itemize}

\section{The $\rho$ parameter and the custodial symmetry}
In this section we comment on the relevance of the $\rho$ parameter
and the custodial symmetry for the study of the $\sbs$ of the
Electroweak Theory.

The $\rho$ parameter is defined as the ratio of neutral to charged
current interactions at low energies:
\begin{equation}
\rho \equiv \frac{T_{NC}(q^2<<M_Z^2)}{T_{CC}(q^2<<M_W^2)}
\end{equation}
and is known from $\nu$-scattering experiments to be very close to one:
$\rho_{\rm exp}\approx 1$.

The SM prediction at tree level is given by:
\begin{equation}
\rho_{\rm tree}^{\rm SM}= \frac{M_W^2}{M_Z^2c^2}
\end{equation}
From this equation and by using the tree level expressions of
eq.(\ref{tree}) one gets the well known result,
\[ \rho_{\rm tree}^{\rm SM}= 1 \]
At one loop and by keeping just the so-called 'oblique' corrections,
namely, the self-energies but not the vertex and box corrections, one
gets:
\begin{equation}
\rho =\frac{\rho_{\rm tree}}{1-\Delta \rho}
\end{equation}
where $\Delta \rho$ can be written in terms of the renormalized
gauge bosons
self-energies as follows:
\begin{equation}
\Delta \rho  =	\frac{\Sigma^{\rm R}_Z (0)}{M_Z^2}
-  \frac{\Sigma^{\rm R}_W (0)}{M_W^2}
\end{equation}
or, alternatively, in terms of the unrenormalized self-energies:
\begin{equation}
\Delta \rho  =	\frac{\Sigma_Z (0)}{M_Z^2}
-  \frac{\Sigma_W (0)}{M_W^2}
- \frac{2 s}{c} \frac{\Sigma_{\gamma Z} (0)}{M_Z^2}
\end{equation}
where, $M_W$, $M_Z$, $s$ and $c$ are the renormalized parameters in the
on-shell scheme \cite{holl}.

In the SM, $\Delta \rho$ receives contributions from both the bosonic
and
fermionic loops and they are of electroweak strength, meaning $\Delta
\rho \sim O(10^{-2})$.

The fact that $\rho$ is very close to one can be understood from the
theory point of view as a consequence of an additional approximate
global symmetry of $\cl_{\rm SM}$. This is the named custodial symmetry,
$SU(2)_C$ \cite{custo} and it would be an exact symmetry if the masses
of the fermions
in each fermionic weak doublet (or, equivalently, their corresponding
Yukawa couplings) were equal, $m_{f_1}=m_{f_2}$, and if
$g'=0$. The existence of mass splittings and interactions mediated by
the hypercharge boson produce some explicit $SU(2)_C$ breaking terms.
However, they are in general  small except for the top-bottom mass
splitting effect.

It is interesting to realize that if one isolates the pure scalar
sector of the SM it turns out to be exactly $SU(2)_C$ symmetric. In
order to
show this let us write the Lagrangian of the scalar sector,
eq.(\ref{lsbs}) in terms of a different parametrization:
\begin{eqnarray}
\cl_{\rm SBS}&=&\frac{1}{4} {\rm Tr}
\left [ (\partial_{\mu}M)^\dagger(\partial^{\mu}M)\right ] -V(M)\;; \\
V(M)&=&\frac{1}{4}\lambda \left [\frac{1}{2} {\rm Tr}(M^\dagger
M)+\frac{\mu^2}{\lambda}\right ]^2
\end{eqnarray}
where $M$ is a $2\times 2$ matrix containing the four real scalar
fields of $\Phi$:
\begin{eqnarray}
M &\equiv&
\sqrt{2}(\widetilde{\Phi}\Phi)=\sqrt{2}\left(\begin{array}{ll}
\phi_0^*&\phi^{\dagger}\\-\phi^-& \phi_0\end{array}\right)
\;;\; \nonumber  \\
\Phi & = & \left(
\begin{array}{c}\phi^+ \\
 \phi_0 \end{array}\right)\;;\;    \nonumber \\
\widetilde{\Phi}&=&i\tau_2\Phi^*=\left(\begin{array}{c}
\phi_0^*\\-\phi^-\end{array}\right)
\end{eqnarray}
It is easy to check that $\cl_{\rm SBS}$ is invariant under the
transformations:
\[M\rightarrow g_L M g_R^+ \;;\;\;g_L\in SU(2)_L\;;\;\;g_R\in SU(2)_R\]

This global symmetry $\lr$ is called the 'chiral' symmetry of the scalar
sector of the SM because of its analogy with the chiral symmetry of QCD.
Furthermore, if one studies the vacuum state of the theory defined by
this pure scalar sector one finds out that it is not 'chiral' invariant
but just invariant under a lower symmetry given by the diagonal subgroup
$\lpr$ which is identified with the custodial symmetry group, $SU(2)_C$.
This is the analogous to the isospin symmetry group of QCD. In summary,
the $\sbs$ of the SM has a spontaneouly broken global chiral symmetry:
\[ \lr \rightarrow SU(2)_C \]

Once the subgroup $\gs$ is gauged, the complete $\cl_{\rm SM}$ is not
anymore $\lr$ nor custodial invariant. The explicit breaking in the
bosonic sector is produced by the hypercharge boson mediated
interactions and
is small since it is proportional to $g'$. Some observables as $\Delta
\rho$ meassure precisely these custodial breaking terms and, therefore,
they vanish
in the limit $g'\approx 0$, that is
$\Delta\rho_{\rm bosonic}\approx 0$.
One crucial point is that this result is
true even in the hypothetical case that the $\sbs$ be strongly
interacting. The prediction of $\rho \approx 1$, at tree level and
beyond tree level, is protected against potentially large corrections
from this sector due to the approximate custodial symmetry of the
Electroweak Theory. It is, therefore, a reasonable choice in model
building beyond the SM to assume that this custodial symmetry is indeed
a symmetry of its corresponding $\sbs$.

\section{Experimental bounds on $M_H$}
The search of the Higgs particle at present $e^+e^-$ and
$\bar{p}p$ colliders is very difficult due the smallness of the
cross-sections for Higgs production \cite{hhg} which, in turn, is
explained
in terms of the small couplings of the Higgs particle to light fermions:
$H\bar{f}f\leftrightarrow -i\frac{g}{2}\frac{m_f}{M_W}$. On the other
hand, at present available energies, the dominant decay channel is
$H\rightarrow b\bar{b}$ (see Fig.3) which is not easy to study due to
the complexity of the final state and the presence of large backgrounds.

\vspace{10 cm}
\begin{center}
{\bf Fig.3} Higgs decay branching ratios (ref.\cite{lhc})
\end{center}
\vspace{0.5cm}

{\bf I.- Higgs search at $e^+e^-$ colliders (LEP, SLC)}

The Higgs search at present $e^+e^-$ colliders is done mainly by
analysing the process \cite{sop,olch}:
\[ e^+e^- \rightarrow Z \rightarrow Z^*H\]
with the virtual $Z^*$ decaying as $Z^*\rightarrow l^+l^-$ or
$Z^*\rightarrow \nu\bar{\nu}$ and the Higgs particle decaying as
$H\rightarrow b\bar{b}$.

At LEPI with a center-of-mass-energy adjusted to the $Z$ mass,
$\sqrt{s}\sim
M_Z$, a very high statistic has been reached and a systematic
search of the Higgs particle for all kinematically allowed $M_H$ values
has been possible. The absence of any experimental signal from the Higgs
particle implies a lower bound on $M_H$. The most recent reported bound
from LEP is \cite{mhexp,olch,sop}:
\[ M_H > 65.1\; GeV \; (95\% C.L.) \]
In the second phase of LEP, LEPII, a center-of-mass-energy of up to
$\sqrt{s}\sim 175 GeV$ is expected to be reached. The relevant
process for Higgs searches will be:
\[ e^+e^-\rightarrow Z^* \rightarrow ZH\]
where now the intermediate $Z$ boson is virtual and the final $Z$ is on
its mass shell. The analyses of the various relevant $Z$ and $H$ decays
will explore the following mass values:
\[ M_H< \sqrt{s}-M_Z \sim 80\; GeV  \]
In addition to the direct bounds on $M_H$ from LEP data, a great effort
is being done also in the search of indirect Higgs signals from its
contribution to electroweak quantum corrections. In fact, there have
been already the first attempts to extract experimental bounds on $M_H$
from the
meassurement of observables as $\Delta \rho$ whose prediction in the SM
is well known. It is interesting to mention that neither the Higgs
particle nor the top quark decouple from these low energy observables.
It means that the quantum effects of a virtual $H$ or $t$ do not vanish
in the limit of infinitely large $M_H$ or $m_t$ respectively. For
instance, the leading corrections to $\Delta \rho$ in
these limits are \cite{screen, cfh}:
\begin{eqnarray}
(\Delta\rho)_t &=&\frac{g^2}{64\pi^2}N_C \frac{m_t^2}{M_W^2}+...
\nonumber \\
(\Delta\rho)_H&=&-\frac{g^2}{64\pi^2}3
\tan^2\theta_W\log\frac{M_H^2}{M_W^2}+...
\end{eqnarray}
Whereas the top corrections grow with the mass as $m_t^2$, the Higgs
corrections are milder growing as $\log M_H^2$. It means that the top
non-decoupling effects at LEP are important. In fact they have been
crucial in the search of the top quark and have provided one of the
first indirect indications of the 'preference of data' for large $m_t$
values. This has been finally confirmed with the discovery of the
top quark at
TEVATRON and the meassurement of its mass \cite{top}, resulting in a
weighted average value of
$m_t=180\pm 12\;GeV$.

The fact that the Higgs non-decoupling effects are soft was announced a
long time ago by T.Veltman in the so-called {\it Screening Theorem}
\cite{screen}. This theorem states that, {\it at one-loop, the dominant
quantum
corrections from a heavy Higgs particle to electroweak observables grow,
at most, as} $\log M_H$. {\it The Higgs corrections are of the generic
form}:
\[g^2(\log\frac{M_H^2}{M_W^2}+g^2\frac{M_H^2}{M_W^2}+...)\]
{\it and the potentially large effects proportional to} $M_H^2$ {\it are
'screened' by additional small} $g^2$ {\it factors}.

Althougth some analyses performed at LEP show a slight 'preference of
data' for a light Higgs ($M_H<600\;GeV,\;95\%C.L.$, with $m_t$ fixed to
the TEVATRON value \cite{olch}) there are still many uncertainties in
this
interpretation and the conclusion is highly dependent on the assumed
input values of $m_t$, $\alpha_s(M_Z^2)$ and $\alpha(M_Z^2)$. In fact,
there are some parallel works \cite{con} where the opposite preference
for
a heavy Higgs is claimed. At present, it is therefore too premature to
reach a definite conclusion and we should wait till the uncertainties in
the input parameters be considerably reduced.

{\bf II.- Higgs search at hadronic colliders }

The relevant subprocesses for Higgs production at hadronic $pp$ and
$p\bar{p}$ colliders are shown in Fig.4.

\vspace{7 cm}
\begin{center}
{\bf Fig.4} Higgs production mechanisms at hadronic colliders
\end{center}
\vspace{0.5cm}

At present available energies the dominant subprocess is $gg$-fusion.
This can be seen in Fig.5 where the cross section for the various Higgs
production channels at the present collider TEVATRON with $\sqrt{s}=2\;
TeV$ are shown.
\newpage
${}^{}$
\vspace{10cm}
\begin{center}
{\bf Fig.5} Higgs production rates at TEVATRON (ref.\cite{htev})
\end{center}
\vspace{0.5cm}

Unfortunately, TEVATRON is not an efficient experiment for Higgs
searches,
mainly because of lack of statistics, even for the case of a light Higgs
where the cross section is maximum. For instance, for the energy and
luminosity values of $\sqrt{s}=1.8\;TeV$ and $\cl
=10^{-31}cm^{-2}sec^{-1}$ a
Higgs particle with $M_H= 60\;GeV$ would produce just about $400$ events
per year. Furthermore, in order to detect a light Higgs at hadron
colliders one must look at the cleanest decays as, for instance, the
$H\rightarrow
\gamma\gamma$ channel which has a too small branching ratio,
$BR(H\rightarrow \gamma\gamma)\sim 10^{-3}$.

The Higgs search at the recently approved LHC collider being built at
CERN, is fortunately more promising \cite{pauss}.
The cross section of Higgs production in
the various modes at LHC are shown in Fig.6.

\newpage
${}^{}$
\vspace{10cm}
\begin{center}
{\bf Fig.6} Higgs production rates at LHC (ref.\cite{lhc})
\end{center}
\vspace{0.5cm}

For $M_H<800\;GeV$ and $m_t\sim 180\;GeV$ the dominant channel is still
$gg$-fusion. However, the $WW$ and $ZZ$ fusion channels become also
relevant in the large $M_H$ region and, in particular, they can provide
valuable information on the Higgs system if it is strongly interacting.

The various exhaustive studies done so far indicate that if the LHC
nominal parameters of $\sqrt{s}=14\; TeV$ and one-year-integrated
luminosity of $L=10^5\;pb^{-1}$ are reached, the whole missing mass
range of $80\;GeV\leq M_H \leq 1\; TeV$ can be covered \cite{lhc}.

\section{$WW$ scattering and the Effective $W$
approximation}
For a very heavy Higgs particle with $M_H\sim O(1\;TeV)$ the Higgs width
is comparable with its mass and to consider vector boson fusion channels
as in Fig.4 is not anymore a good aproximation. In this case, all
diagrams that participate in the $VV$ scattering subprocess
($V=W^{\pm},Z$) must be included. For instance, the diagrams
contributing to $W^+W^-$ production from $W^+W^-$ fusion are shown in
Fig.7.

\newpage
${}^{}$
\vspace{13cm}
\begin{center}
{\bf Fig.7} Contributing diagrams to  $qq\rightarrow qqWW$
through $WW$ scattering
\end{center}
\vspace{0.5cm}

The computation of all these diagrams is quite lengthy. For
simplicity, it is convenient to use the so-called Effective $W$
Approximation \cite{weff} which is very similar to the well known
Effective Photon
Approximation \cite{pheff}. In the $W$ Effective
Approximation, the process $qq\rightarrow qqVV$ with $V=W^{\pm},Z$
is factorized out into the production of 'quasireal' $V$'s being
radiated from the initial quarks and the posterior rescattering of these
bosons which are assumed to be on-shell. It works well because the
cross section of the full process $qq\rightarrow qqVV$ is
known to be dominated by the kinematical region where the intermediate
$V$'s in this process are emited close to their mass shell and with
small scattering angles with respect to the outgoing quark. From the
computational point of view, the $V$'s are considered as partons with
certain probability distributions and the cross section of the full
process is obtained by making the convolution of these functions with
the cross section of the corresponding $VV$ scattering subprocess. For
instance, for the $WW$ case one writes:
\begin{equation}
\sigma(pp\rightarrow (WW\rightarrow WW)X)=\int_{\tau_{\rm min}}^1
d\tau \left ( \frac{d\cl}{d\tau} \right )_{pp/WW}\;\sigma(WW\rightarrow
WW)
\end{equation}
where the luminosity of $W$'s from the
protons is given by
\begin{equation}
\left ( \frac{d\cl}{d\tau}\right )_{pp/WW}
=\sum_{ij}\int_{\tau}^1
\frac{d\tau'}{\tau'}\int_{\tau'}^1\frac{dx}{x}q_i(x)
q_j(\frac{\tau'}{x})
\left ( \frac{d\cl}{d\zeta}\right )_{q_iq_j/WW}
\end{equation}
and the corresponding one from the quarks is
\begin{equation}
 \left (
\frac{d\cl}{d\zeta}\right )_{q_iq_j/WW}=\int_{\zeta}^1\frac{dy}{y}
f_{q/W}(y)f_{q/W}(\frac{\zeta}{y})\;;\;\zeta=\frac{\tau}{\tau'}
\end{equation}
Here, $q_i(x)$ are the quark distribution functions in the proton and
$f_{q/W}$ is the $W$ distribution function in the quark $q$. The
simplest
version of these functions for longitudinal and tranverse $W$'s are the
following,
\begin{equation}
f^L_{q/W}=\frac{g^2}{16\pi^2}\left ( \frac{1-x}{x}\right ) \;;\;
f^T_{q/W}(x)=\frac{g^2}{64\pi^2}\log \left ( \frac{4E^2}{M_W^2}\right )
\left ( \frac{1+(1-x)^2}{x} \right )
\end{equation}
Here $x$ is the momentum fraction of the quark $q$ that is carried by
the emited $W$ and $2E$ is the total energy of the $qq$ system. We
have averaged over the two transverse polarizations.
 Similar equations are provided for the case of $Z$ gauge bosons.

The Effective $W$ Approximation is particularly useful in the case where
the $\sbs$ is strongly interacting and more generaly for the kind of
models that predict  different $V_LV_L\rightarrow V_LV_L$ scattering
amplitudes than those of the SM.

\section{The Equivalence Theorem}

This theorem states the following \cite{et}:

{\it The scattering amplitudes of longitudinal gauge bosons} $V_L$
($V=W^{\pm},Z$), {\it at high energies}, $\sqrt{s}>>M_V$, {\it are
equivalent
to the scattering amplitudes of their corresponding would-be Goldstone
bosons} $w$,
\begin{equation}
T(V_L^1V_L^2...V_L^N \rightarrow V_L^1V_L^2...V_L^{N'})\approx
i^N(-i)^{N'}T(w_1w_2...w_N\rightarrow w_1w_2...w_{N'})
\label{et}
\end{equation}
It can be seen as the reflect that in a gauge theory with spontaneous
symmetry breaking, the needed longitudinal polarization degrees  of
the massive vector bosons are originated by the Higgs Mechanism
precisely from the corresponding would-be Goldstone bosons.

The Equivalence Theorem works at tree level and beyond tree level and it
is very useful in simplifying a number of involved computations.
For instance, it can be applied to compute the partial widths of a heavy
Higgs into longitudinal $W$ and $Z$ bosons. In this case,
\begin{eqnarray}
\Gamma(H\rightarrow W_L^+W_L^-)&=&\Gamma(H\rightarrow
w^+w^-)+O(\frac{M_W}{M_H})\nonumber \\
\Gamma(H\rightarrow Z_LZ_L)&=&\Gamma(H\rightarrow
zz)+O(\frac{M_Z}{M_H})
\end{eqnarray}
where $w^{\pm}$ and $z$ are the three would-be Goldstone bosons of the
Electroweak Theory. At tree level it gives,
\begin{equation}
\Gamma(H\rightarrow W_L^+W_L^-) \simeq
2\Gamma(H\rightarrow Z_LZ_L) \simeq \frac{g^2}{64\pi}
\frac{M_H^3}{M_W^2}
\end{equation}
It has been computed also to one loop \cite{will} and recently to two
loops \cite{ghin}.

One of the most important applications of the Equivalence Theorem is in
the study  of $VV$ scattering at the future LHC collider. Let us see,
for instance, how it works in the case of
$W_L^+W_L^- \rightarrow W_L^+W_L^-$
scattering at tree level.

The contributing diagrams, in the SM, to the amplitudes with external
$W_L$'s are the same as in the ring of Fig.7.  The diagrams for
scattering with external $w$'s are shown in Fig.8.

\vspace{7 cm}
\begin{center}
{\bf Fig.8} Contributing diagrams to  $w^+w^-\rightarrow w^+w^-$
scattering
\end{center}
\vspace{0.5cm}

Here the polarization vector of an external $W_L$ with momentum $k$ is
given by  \[ \epsilon_L^{\mu}=\frac{1}{M_W}(|\vec{k}|,0,0,k_0) \]

After the computation of these diagrams one gets the following results:
\begin{eqnarray}
T(W_L^+W_L^-\rightarrow W_L^+W_L^-)&=&
-\frac{1}{v^2}\{-s-t+\frac{s^2}{s-M_H^2}+\frac{t^2}{t-M_H^2}
      +2M_Z^2+
\nonumber \\ & &
      \frac{2M_Z^2s}{t-M_Z^2}+
\frac{2t}{s}(M_Z^2-4M_W^2)-\frac{8s_W^2M_W^2M_Z^2s}{t(t-M_Z^2)}\}
\label{long}  \\
T(w^+w^-\rightarrow w^+w^-)&=&
-\frac{M_H^2}{v^2}\{\frac{s}{s-M_H^2}+
\frac{t}{t-M_H^2}\}
\end{eqnarray}
By studying the above expressions in the large energy limit,
$\sqrt{s}>>M_W,M_Z$, and by keeping just the leading term, one can
check the validity of the Equivalence Theorem which in this case reads,
\begin{equation}
T(W_L^+W_L^-\rightarrow W_L^+W_L^-)=
T(w^+w^-\rightarrow w^+w^-)+O(\frac{M^2}{s})
\end{equation}
and it holds for any value of $M_H$. The above amplitudes have been
computed up to one loop \cite{dwvy} and the Equivalence Theorem seems to
work well.

\section{Theoretical bounds on $M_H$}
In this section we summarize the present bounds on $M_H$ from the
requirement of consistency of the theory.

{\bf I.- Upper bound on $M_H$ from Unitarity}

\vspace{0.5 cm}
Unitarity of the scattering matrix together with the elastic
approximation for the total cross-section and the Optical Theorem
imply certain elastic unitarity conditions for the partial wave
amplitudes. These, in turn, when applied in the SM to scattering
processes involving the Higgs particle, imply an upper limit on the
Higgs mass. Let us see this in more detail for the simplest case of
scattering of massless scalar particles: $1+2\rightarrow 1+2$.

The decomposition  of the amplitude in terms of partial waves is given
by:
\begin{equation}
T(s,\cos\theta)=16\pi\sum_{J=0}^{\infty}(2J+1)a_J(s)P_J(\cos\theta)
\end{equation}
where $P_J$ are the Legendre polynomials.

The corresponding differential cross-section is given by:
\begin{equation}
\frac{d\sigma}{d\Omega}=\frac{1}{64\pi^2s}|T|^2
\end{equation}

Thus, the elastic cross-section is written in terms of partial
waves as:
\begin{equation}
\sigma_{\rm el}=\frac{16\pi}{s}\sum_{J=0}^{\infty}(2J+1)|a_J(s)|^2
\label{elast}
\end{equation}
On the other hand, the Optical Theorem relates the total cross-section
with the forward elastic scattering amplitude:
\begin{equation}
\sigma_{\rm tot}(1+2\rightarrow {\rm anything})=\frac{1}{s}
 Im\;T(s,\cos\theta=1)
\label{tot}
\end{equation}
In the elastic approximation for $\sigma_{\rm tot}$ one gets
$\sigma_{\rm tot}\approx \sigma_{\rm el}$.
From this and by identifying
eqs.(\ref{elast}) and (\ref{tot}) one finally finds,
\begin{equation}
Im\;a_J(s)=|a_J(s)|^2\;;\;\forall J
\label{part}
\end{equation}
This is called the elastic unitariry condition for partial wave
amplitudes. It is easy to get from this eq.(\ref{part}) the following
inequalities:
\begin{equation}
|a_J|^2\leq 1\;;\;0\leq Im\;a_J\leq 1\;;\;|Re\; a_J|\leq \frac{1}{2}
\;;\; \forall J
\end{equation}
These are necessary but not sufficient conditions for elastic unitarity.
It implies that if any of them is not fulfiled then the elastic
unitarity
condition of eq.(\ref{part}) also fails, in which case the unitarity of
the theory is said to be violated.

Let us now study the particular case of $W_L^+W_L^-$ scattering in the
SM and find its unitarity conditions. The $J=0$ partial wave can be
computed from:
\begin{equation}
a_0(W_L^+W_L^-\rightarrow W_L^+W_L^-)=\frac{1}{32\pi}\int_{-1}^1
T(s,\cos\theta)d(\cos\theta)
\end{equation}
wher the amplitude $T(s,\cos \theta)$ is given in eq.(\ref{long}). By
studying the large energy limit  of
$a_0$ one finds,
\begin{equation}
|a_0| \stackrel{s>>M_H^2,M_V^2}{\longrightarrow} \frac{M_H^2}{8\pi v^2}
\end{equation}
Finally, by imposing the unitarity condition $|Re\;a_0|\leq \frac{1}{2}$
one gets the following upper bound on the Higgs
mass:
\begin{equation}
M_H<860\; GeV
\end{equation}
One can repeat the same reasoning for different channels and find
similar or even tighter bounds than this one.

At this point, it should be mentioned that these upper bounds based on
perturbative unitarity do not mean that the Higgs particle cannot be
heavier than these values. The conclusion should be, instead, that for
those large $M_H$ values a perturbative approach is not valid and
non-perturbative techniques are required.

{\bf II.- Upper bound on $M_H$ from Triviality}

\vspace{0.5 cm}
Triviality in $\lambda \Phi^4$ theories \cite{triv} (as, for instance,
the scalar
sector of the SM) means that the particular value of the renormalized
coupling of $\lambda_R=0$ is the unique fixed point of the theory. A
theory with
$\lambda_R=0$
contains non-interacting particles and therefore it is trivial. This
behaviour can already be seen in the renormalized coupling at one-loop
level:
\begin{equation}
\lambda_R(Q)=\frac{\lambda_0}{1-\frac{3}{2\pi^2}\lambda_0
\log(\frac{Q}{\Lambda})}\;;\;\lambda_0\equiv \lambda_R(Q=\Lambda)
\end{equation}
As we attempt to remove the cut-off $\Lambda$ by taking the limit
$\Lambda \rightarrow \infty$ while $\lambda_0$ is kept fixed to an
arbitrary but finite value, we find out that
$\lambda_R(Q)\rightarrow 0$ at any finite energy value $Q$. This, on the
other hand, can be seen as a consequence of the existence of the well
known Landau pole of $\lambda \Phi^4$ theories.

The trivilaty of the $\sbs$ of the SM is cumbersome since we need a
self-interacting scalar system to generate $M_W$ and $M_Z$ by the Higgs
Mechanism. The way out from this apparent problem is to assume that the
Higgs potential $V(\Phi)$ is valid just below certain 'physical' cut-off
$\Lambda_{\rm phys}$. Then, $V(\Phi)$ describes an effective low energy
theory which emerges from some (so far unknown) fundamental physics with
$\Lambda_{\rm phys}$ being its characteristics energy scale. We are
going to see next that this assumption implies an upper bound on $M_H$
\cite{1loop}.

Let us assume some concrete renormalization of the SM parameters. The
conclusion does not depend on this particular choice. Let us define, for
instance, the renormalized Higgs mass parameter as:
\begin{equation}
M_H^2 = 2 \lambda_R(v)v^2
\label{mh}
\end{equation}
where,
\begin{equation}
\lambda_R(v)=\frac{\lambda_0}{1-\frac{3}{2\pi^2}\lambda_0
\log(\frac{v}{\Lambda_{\rm phys}})}
\label{lr}
\end{equation}
Now, if we want $V(\Phi)$ to be a sensible effective theory, we must
keep all the renormalized masses below the cut-off and, in particular,
$M_H<\Lambda_{\rm phys}$. However, from eqs.(\ref{mh}) and (\ref{lr})
we can see that for arbitrary values of $\Lambda_{\rm phys}$ it is not
always possible. By increasing the value of $\Lambda_{\rm phys}$, $M_H$
decreases and the other way around, by lowering $\Lambda_{\rm phys}$,
$M_H$ grows. There is a crossing point where $M_H\approx \Lambda_{\rm
phys}$ which happens to be around an energy scale of approximately
$1\;TeV$. Since we want to keep the Higgs mass below the physical
cut-off, it implies finally the announced upper bound,
\[M_H^{\rm 1-loop}< 1\;TeV\]
Of course, this should be taken just as a perturbative estimate of the
true triviality bound. A more realistic limit must come from a
non-perturbative treatment. In particular, the analyses performed on
the lattice \cite{lattice} confirm this behaviour and place even tighter
limits. The following bound is found,
\[ M_H^{\rm Lattice} < 640 \;GeV \]
Finally, a different but related perturbative upper limit on $M_H$ can
be found by analysing the renormalization group equations in the SM to
one-loop. Here one includes, the scalar sector, the gauge boson
sector and restricts the fermionic sector to the third generation.  By
requiring the theory to be perturbative (i.e. all the couplings be
sufficiently small) at all the energy scales below some fixed high
energy, one finds a maximum allowed $M_H$ value \cite{cab}.
For instance, by fixing
this energy scale to $10^{16}\;GeV$ and for $m_t=170\;GeV$ one gets:
\[ M_H^{\rm RGE} < 170\;GeV \]
Of course to believe in perturbativity up to very high energies
could be just a theoretical prejudice. The existence
of a non-perturbative regime for the scalar sector of the SM is still a
possibility and one should be open to new proposals in this concern.

{\bf III.- Lower bound on $M_H$ from Vacuum Stability}

\vspace{0.5 cm}
Once the asymmetric vacuum of the $\gs$ theory has been fixed, one must
require
this vacuum to be stable under quantum corrections. In principle,
quantum corrections could destabilize the asymmetric vacuum and change
it to the symmetric one where the spontaneous symmetry breaking does
not take place. This phenomenom can be better explained in terms of the
effective potential with quantum corrections included on it. Let us
take, for instance, the effective potential of the Electroweak Theory
to one loop in the small $\lambda$ limit:
\begin{equation}
V_{\rm eff}^{\rm 1-loop}(\Phi) \simeq -\mu^2\Phi^{\dagger}\Phi+
\lambda_R(Q_0) (\Phi^{\dagger}\Phi)^2+\beta_{\lambda}
(\Phi^+\Phi)^2\log\left ( \frac{\Phi^{\dagger}\Phi}{Q_0^2} \right )
\end{equation}
where, $\beta_{\lambda} \equiv \frac{d\lambda}{dt} \simeq
 \frac{1}{16\pi^2}\left [
-3\lambda_t^4+\frac{3}{16}(2g^4+(g^2+g'^2)^2)\right ]$.

The condition of extremum is:
\begin{equation}
\frac{\delta V_{\rm eff}^{\rm 1-loop}}{\delta\Phi}=0
\end{equation}
which leads to two possible solutions: a) The trivial vacuum with
$\Phi=0$; and b) The non-trivial vacuum with $\Phi=\Phi_{\rm vac}\neq
0$. If we want the true vacuum to be the non-trivial one we
must have:
\begin{equation}
V_{\rm eff}^{\rm 1-loop}(\Phi_{\rm vac}) <
V_{\rm eff}^{\rm 1-loop}(0)
\label{stab}
\end{equation}
However, the value of the potential at the minimum depends on the size
of its second derivative:
\begin{equation}
M_H^2\equiv \frac{1}{2}\{\frac{\delta^2V}
{\delta\Phi^2} \}_{\Phi=\Phi_{\rm vac}}
\end{equation}
and, it turns out that for too low values of $M_H^2$ the condition
above, eq.(\ref{stab}), turns over. That is, $V(0)<V(\Phi_{\rm vac})$
and the true vacuum changes to the trivial one. The condition for vacuum
stability then implies a lower bound on $M_H$ \cite{vstab1}. More
precisely,
\begin{equation}
M_H^2>\frac{3}{16\pi^2v^2}(2M_W^4+M_Z^4-4m_t^4)
\end{equation}
Surprisingly, for $m_t>78\;GeV$ this bound dissapears and, moreover,
$V_{\rm eff}^{\rm 1-loop}$
becomes unbounded from below!. Apparently it seems a
disaster since the top mass is known at present and is certainly larger
than this value. The solution to this problem relies in the fact that
for such input values, the 1-loop approach becomes unrealistic and a
2-loop analysis of the effective potential is needed. Recent studies
indicate that by requiring vacuum stability at 2-loop level and up to
very large energies of the order of $10^{16}\;GeV$, the following lower
bound is found \cite{vstab2}:
\begin{equation}
M_H^{\rm v.stab.}> 132\;GeV
\end{equation}
This is for $m_t=170\; GeV$ and $\alpha_s=0.117$ and there is an
uncertainty in this bound of $5$ to $10$ $GeV$ from the uncertainty in
the $m_t$ and $\alpha_s$ values.

\section{The naturalness Problem}
In order to show the so-called naturalness problem, let us compute
first the
renormalized Higgs mass, $M_H^R$, to one-loop in the SM. Since the SM is
a renormalizable theory we can apply the renormalization program as
usual.
Let us choose, for instance, the on-shell scheme where $M_H^R$ coincides
with the physical mass $M_H$ and it is related with the bare
(unphysical) mass
$M_H^0$ by:
\begin{eqnarray}
(M_H^R)^2&=&(M_H^0)^2+\delta M_H^2 \nonumber \\
\delta M_H^2&=&{\rm Re}\;\Sigma_{H}\left[ (M_H^R)^2\right]
\end{eqnarray}
where $-i\Sigma_{H}\left[ q^2\right ] $ is given by the sum of
the 1-loop diagrams contributing to the Higgs self-energy. Some of these
diagrams are shown in Fig.9.

\vspace{6 cm}
\begin{center}
{\bf Fig.9} Some contributing diagrams to  the Higgs  self-energy
\end{center}
\vspace{0.5cm}

Some of these diagrams are quadratically divergent and
some others are logarithmically divergent. This can be
easily shown by computing the integrals  with a cut-off $\Lambda$ in the
ultraviolet region. For instance,
\begin{eqnarray}
\left[ -i\Sigma_{H}^{\rm a}\right]_{\rm div}& \sim &
\frac{3}{16\pi^2}\lambda_R \Lambda^2 \nonumber \\
\left[ -i\Sigma_{H}^{\rm b}\right]_{\rm div}& \sim &
\frac{\lambda_R^2 v^2}{16 \pi^2}\log (\frac{\Lambda}{M_W})
\end{eqnarray}
The relation between the renormalized and the bare masses is,
therefore, of the following generic form at on-loop:
\begin{equation}
(M_H^R)^2=(M_H^0(\Lambda))^2+\left[C_1\Lambda^2+C_2\log\Lambda+C_3\right]
\;;\; C_i=O(\frac{1}{16\pi^2})
\label{mhr}
\end{equation}
The renormalization program tells us how the unphysical
$M_H^0(\Lambda)$ must be fixed in order to absorve all the divergences
and to get a finite $M_H^R$ in the $\Lambda \rightarrow \infty$ limit.
The cut-off here is unphysical and must be removed at the end from
physical
quantities. So far, there is nothing unnatural. It is just the standard
renormalization procedure.

The problem arises if (and only if) one wants to interpret the SM as a
low energy effective theory of some fundamental theory which operates
at very high energies, say $M_{\rm GUT}\sim 10^{16}\;GeV$ or
$M_{\rm Planck}\sim 10^{19}\;GeV$, etc.. In this case, the cut-off
becomes
a physical quantity and must be related somehow to the energy at which
the new physics of the fundamental theory manifests. Besides,
$M_H^0(\Lambda)$ must be predictable from this underlying theory. For
high cut-off
values, the 1-loop corrections in eq.(\ref{mhr}) tend to push $M_H^R$
to high values as well, so that if one wishes to keep $M_H^R$ within a
resonable low energy range, $M_H^0(\Lambda)$ must be adjusted
accordingly. Sometimes this adjustment is critical.
For instance, for $\Lambda=M_{\rm Planck}$ and if we
require $M_H^R<1\;TeV$, $M_H^0$ must be fine-tuned up to 30 decimals!.
This extreme fine-tuning is what is considered unnatural.

There are two most popular proposed solutions to the naturalness
problem, one
is based on Supersymmetry and the other one on Technicolor.

{\bf I.- Supersymmetry}

The SM is assumed to be the low energy effective theory of some
fundamental theory which is supersymmetric and operates at very high
energy, say $\Lambda_{\rm SUSY}\sim O(M_{\rm Planck})$.

The new symmetry between bosons and fermions, the Supersymmetry (SUSY),
\cite{susy1} implies
an extension  of the SM spectrum. In particular, for each scalar boson
particle with mass $m_{\rm particle}$ there must exist a fermionic
superpartner with its same mass, $m_{\rm sparticle}=m_{\rm particle}$.
Thus, if this Supersymmetry is exact there is an exact cancelation
between each 1-loop diagram with a scalar particle flowing in the loop
and the corresponding diagram with its fermionic superpartner in the
loop.
As a consequence the quadratic divergences vanish
 and only softer logarithmic divergences remain:
\begin{equation}
(M_H^R)^2=(M_H^0(\Lambda_{\rm SUSY}))^2+\left[
\hat{C_2}\log\Lambda_{\rm SUSY}+\hat{C_3}\right]
\end{equation}
Therefore, no unnatural fine-tuning is needed in the exact SUSY case.

However, the absence of scalar particles in the spectrum with the same
mass as the known fermions  indicates that the Supersymmetry must be
broken at low energies (i.e, energies available at present experiments).
If SUSY is not exact the particles and their superpartners are not
degenerated anymore and $m_{\rm sparticle}$ must be larger than
$m_{\rm particle}$. However, it cannot be too large if we want the
naturalness problem not to come back. That is to say, if $m_{\rm
sparticle}> m_{\rm particle}$ the quadratic divergences reappear and
contribute to $(M_H^R)^2$ as:
\[ (m_{\rm sparticle}^2-m_{\rm particle}^2 )\frac
{\Lambda_{\rm SUSY}^2}{m_{\rm sparticle}^2} \]
Therefore, to keep the fine-tuning controled at less than a few percent
level, the sparticle spectrum must appear below about $1\;TeV$:
\[ m_{\rm sparticle}\leq O(1\;TeV) \]
It announces an interesting phenomenology for sparticle searches at
present and future colliders \cite{susy2}. In particular the LHC
collider seems very promissing.

{\bf II.- Technicolor}

In this class of theories, the $\sbs$ does not contain an elementary
Higgs particle and the symmetry breaking, $\gs \rightarrow
U(1)_{\rm em}$,
is produced dinamically by new gauge interactions of a $SU(N_{TC})$
gauge theory \cite{tc}. It is a confining theory at large distances and
has strong
interactions similar to $SU(3)_C$ of QCD. Because of this analogy with
QCD, $SU(N_{TC})$ is called Technicolor Theory. The fundamental fields
are the techniquarks $q_{TC}$ and technigluons $G_{TC}$, and $N_{TC}$ is
the total number of technicolors.

The absence of an elementary Higgs boson in Technicolor
theories automatically avoids the naturalness problem. The Higgs
Mechanism is implemented by a techniquark condensate in analogy to the
quark condensate of QCD:
\begin{equation}
<0|\bar{q}_{TC}q_{TC}|0> \neq 0
\label{cond}
\end{equation}
This condensate must have non-vanishing $SU(2)_L$ and $U(1)_Y$ charges
in order to produce the wanted breaking.
On the other hand, the strong interactions of $SU(N_{TC})$
are assumed to be the responsible for producing these condensates.

\section{The Spectrum of Technicolor}
In Technicolor Theory $SU(N_{TC})$ there is an additional
symmetry, $\lr$, and it happens to be the same global symmetry of the
$\sbs$ of the SM. (In more complex technicolor models this symmetry can
be even larger). Moreover, this symmetry is spontaneously broken by the
condensate eq.(\ref{cond}) to the diagonal subgroup:
\[ \lr \rightarrow \lpr \]
By virtue of the Goldstone Theorem, this breaking leads to the existence
of three Goldstone bosons, the so-called technipions $\pi^{\pm}_{TC}$
and $\pi^0_{TC}$. When the subgroup $\gs$ is gauged the Higgs Mechanism
takes place: The three would-be-Goldstone bosons dissapear from the
spectrum and they are replaced by the longitudinal gauge bosons,
$W^{\pm}_L$ and $Z_L$.

The coupling of the technipions to the weak
current is given by:
\begin{equation}
<0|J^{+\mu}_L|\pi^-_{TC}(p)>=\frac{iF_{\pi}^{TC}}{\sqrt{2}}p^{\mu}
\end{equation}
where the technipion decay constant is:
\[ F_{\pi}^{TC}=v=246\;GeV \]
which is obviously the analogous to $f_{\pi}$ of QCD.

The spectrum of Technicolor is a copy of the spectrum of QCD as well:
Technipions $(\pi_{TC}^{\pm},\pi_{TC}^0)$, Technirhos
$(\rho_{TC}^{\pm}, \rho_{TC}^0)$, etc...An estimate of their masses and
widths can be obtained by rescaling the corresponding values in QCD with
an appropriate factor. This factor can be written as:
\[ \frac{F_{\pi}^{TC}}{f_{\pi}}
\cdot f(N_{TC},N_C) \]
where,
\[ \frac{F_{\pi}^{TC}}{f_{\pi}}=\frac{246\;GeV}{0.094\;GeV}\sim 2700 \]
and $f(N_{TC},N_C)$ is a function of the number of technicolors and the
number of colors. A naive estimate of this function can be got by
using the large $N_{TC}$ approximation in $SU(N_{TC})$ and, similarly,
large
$N_C$ in $SU(N_C)$ \cite{largen}. Thus, for instance, by knowing the
behaviour
at large $N_{TC}$ and $N_C$ of the technimeson and meson parameters
respectively,
\begin{eqnarray}
m_{\rm meson}& \sim & O(1)\;;\;\;\; f_{\pi} \sim  \sqrt{N_C} \nonumber
\\ m_{\rm Tmeson}& \sim & O(1)\;;\;\;\; F_{\pi}^{TC}  \sim \sqrt{N_{TC}}
\end{eqnarray}
one finds out,
\begin{equation}
\frac{m_{\rm Tmeson}}{m_{\rm meson}} \sim
\frac{F_{\pi}^{TC}}{f_{\pi}}\cdot \sqrt{\frac{N_C}{N_{TC}}}
\end{equation}
and, therefore, the first expected resonance is the technirho with
a mass of
\begin{equation}
m_{\rho_{TC}}=
\frac{F_{\pi}^{TC}}{f_{\pi}}\cdot \sqrt{\frac{N_C}{N_{TC}}}
m_{\rho}
\end{equation}
For instance, for $N_C=3$, $N_{TC}=4$  and $m_{\rho}=760\;MeV$ one gets,
\[ m_{\rho_{TC}}=1.8\;TeV \]
This, on the other hand, gives the order of magnitude of the effective
cut-off of Technicolor Theory where the new physics sets in:
\[ \Lambda_{TC}^{\rm eff} \sim O(1\;TeV) \]
Similar arguments can be applied to get the technimeson widths,
\begin{eqnarray}
\Gamma_{\rm meson} \sim O(\frac{1}{N_C})&;&
\Gamma_{\rm Tmeson} \sim O(\frac{1}{N_{TC}}) \nonumber \\
\Rightarrow & &\frac{\Gamma_{\rm Tmeson}}{\Gamma_{\rm meson}}\sim
\frac{N_C}{N_{TC}}\frac{m_{\rm Tmeson}}{m_{\rm meson}}\nonumber \\
\Rightarrow & & \Gamma_{\rho_{TC}}=\frac{N_C}{N_{TC}}
\frac{m_{\rho_{TC}}}{m_{\rho}}\Gamma_{\rho}
\end{eqnarray}
For instance, for $N_C=3$, $N_{TC}=4$ and $\Gamma_{\rho}=151\;MeV$ one
gets:
\[ \Gamma_{\rho_{TC}}= 260\; GeV \]
The prediction of spectrum at the $TeV$ energies in Technicolor
Theories opens new possibilities for particle searches at future
colliders as LHC.

At this point, one should mention about the drawbacks of Technicolor
Theories. It is known that in order to generate the fermion masses one
needs more complex models as the so-called Extended Technicolor Theories
\cite{etc}
which are not free of problems. For instance, it is very difficult to
avoid flavor changing neutral currents in these models. There are
various versions of Technicolor Theories dealing with these problems,
but we are not going to discuss them here. For more information, we
address the reader to ref.\cite{tcrev}.

\section{The Higgs Sector in MSSM}

The Minimal Supersymmetric Standard Model is the simplest
extension of the SM, $SU(3)_{\rm C}\times \gs$, that is supersymmetric
and contains the minimal particle spectrum.

The following are some of the assumptions done to build up this model:
\begin{itemize}
\item[-]
The MSSM is considered as a low energy effective theory that should be
used just at energies below the effective scale of Supersymmetry,
$\Lambda_{\rm SUSY}^{\rm eff} \sim O(1\;TeV)$.
\item[-]
The MSSM comes from a more fundamental Supersymmetric Theory (It could
be, Supergravity, Superstring Theory, etc...)
which operates at much higher energy scales, say $\Lambda_{\rm SUSY}\sim
O(M_{\rm Planck})$, and where SUSY is an exact symmetry.
\item[-]
In MSSM it is not needed to specify the particular fundamental theory,
but whatever it might result, it must be the responsible for generating
the so-called soft-SUSY-breaking terms which must be included in the
Lagrangian of MSSM. These terms are needed to break SUSY at low
energies and to explain $m_{\rm sparticle}>m_{\rm particle}$.
\item[-]
The breaking of SUSY in MSSM must be soft to guarantee that no
quadratic
divergences reappear in the scalar selfenergies and, thus, to avoid the
naturalness problem could emerge again.
\item[-]
The soft terms of MSSM are fixed in a way that can produce as
well the wanted breaking: $\gs \rightarrow U(1)_{\rm em}$.
\end{itemize}
In the following we present the Higgs sector in MSSM and tell how the
breaking of $\gs$ occurs. For the remaining spectrum and more
information on MSSM we address the reader to Haber's Lectures in
ref.\cite{susy1} where most of this section has been borrowed from.

In addition to the complex scalar doublet of the SM $\Phi$ with
$Y(\Phi)=1$, we need in SUSY theories a second complex scalar doublet
with opposite hypercharge. Let $H_1$ and $H_2$ be these two doublets:
\begin{eqnarray}
H_1=\left( \begin{array}{c}H_1^0\\H_1^- \end{array} \right )&;&
Y(H_1)=-1 \nonumber \\
H_2=\left( \begin{array}{c}H_2^+\\H_2^0 \end{array} \right )&;&
Y(H_2)=+1
\end{eqnarray}
Because of the Supersymmetry, in addition to these scalars, there are
two associated fermionic superpartners with their same quantum numbers:
\begin{eqnarray}
\widetilde{H_1}=\left( \begin{array}{c}\widetilde{H_1^0}\\
\widetilde{H_1^-}
\end{array}
\right
)&;& Y(\widetilde{H_1})=-1 \nonumber \\
\widetilde{H_2}=\left( \begin{array}{c} \widetilde{H_2^+}\\
\widetilde{H_2^0}
\end{array}
\right
)&;& Y(\widetilde{H_2})=+1
\end{eqnarray}
The reason to introduce two scalar doublets instead of one as in the SM
is two fold: 1) Only by including the fermionic doublets in pairs it is
possible to cancel their contribution to the gauge anomaly; 2) A second
scalar doublet $H_1$ with $Y=-1$ is needed to generate the masses of the
$u$ type quarks. The complex conjugate $H_2^*$ cannot play this role, as
in the SM does, since the requirement of SUSY implies the Superpotential
must be  an analytic function and therefore it does not allow for
complex conjugate scalar fields.

The counting of polarization degrees in the electroweak symmetry
breaking within the MSSM is different than in the SM. It goes as
follows,
\begin{itemize}
\item[1)] {\bf Before SSB} \\
4 massless gauge bosons: $W^{\mu}_{1,2,3}, B^{\mu}$ \\
8 massless scalars: The 4+4 real components of the complex doublets
$H_1$ and $H_2$ \\

Total number of polarization degrees $= 4\times 2 + 8 = 16$
\item[2)] {\bf After SSB} \\
3 massive gauge bosons: $W^{\pm},Z$ \\
1 massless gauge boson: $\gamma$ \\
5 massive scalar bosons: $H^{\pm},A^0,H^0,h^0$ \\

Total number of polarization degrees: $3\times 3+1\times 2+5=16$
\end{itemize}
As in the SM, the total number of polarization degrees is preserved in
the breaking but now it is larger than in SM. Apart from the three
needed
would-be-Goldstone bosons, there have been introduced 'ad hoc' five
more polarization degrees (instead of one as in the SM) which, after the
breaking, emerge as the five physical massive Higgs bosons of the MSSM:
\begin{itemize}
\item[-] two charged scalar bosons, $H^+$ and $H^-$
\item[-] one CP-odd neutral scalar boson, $A^0$
\item[-] two CP-even neutral scalar bosons, $H^0$ and $h^0$
\end{itemize}

The simplest potential im terms of $H_1$ and $H_2$ that is
supersymmetric is given by:
\begin{equation}
V(H_1,H_2)=|\mu|^2 (|H_1|^2+|H_2|^2)+\frac{1}{8}(g^2+g'^2)
(|H_1|^2-|H_2|^2)^2+\frac{1}{2}g^2|H_1^*H_2|^2
\end{equation}
This should be compared with the Higgs potential of the SM:
\begin{equation}
V(\Phi)=- \mu^2\Phi^\dagger \Phi + \lambda (\Phi^\dagger \Phi)^2
\end{equation}
The following observations are in order:
\begin{itemize}
\item[-]
In the SUSY potential there is the same coefficient for the $|H_1|^2$
and $|H_1|^2$ terms. This will lead to some mass relations.
\item[-]
In the SUSY potential the scalar self-coupling is not an
independent papameter but it is given in terms of the gauge coupling
constants $g^2$ and $g'^2$. Therefore, the Higgs sector in the MSSM is
allways weakly interacting and contains, at least, one light Higgs boson
with $m_{h^0} \sim O(100\;GeV)$.
\item[-]
The SUSY potential cannot produce the wanted electroweak
symmetry breaking since it is definite possitive, $V(H_1,H_2)\geq 0
\;\;\;\forall H_{1,2}\;$, and its minimum is at the trivial vacuum,
$H_1=H_2=0$. In order to generate a non-trivial asymmetric vacuum some
additional terms are needed in the potential. These are the above
mentioned soft-breaking terms whose role is two fold: To break the
Supersymmetry and to break $\gs$.
\end{itemize}
The simplest potential with soft breaking terms included is given by,
\begin{eqnarray}
V_{\rm MSSM}&=&m_{1H}^2 |H_1|^2+m_{2H}^2
|H_2|^2-m_{12}^2(\epsilon_{ij}H_1^iH_2^j+h.c.)\nonumber \\
& & +\frac{1}{8}(g^2+g'^2)(|H_1|^2-H_2|^2)^2+\frac{1}{8}g^2|H_1^*H_2|^2
\end{eqnarray}
where,
\begin{equation}
m_{iH}^2 \equiv |\mu|^2+m_i^2\;;\; i=1,2
\end{equation}
and $m_1^2$, $m_2^2$ and $m_{12}^2$ are the soft SUSY breaking
parameters.

One can check that the following are the necessary
conditions for $\gs$ breaking:
\begin{itemize}
\item[1.-] $ |m_{12}^2|^2 >m_{1H}^2 m_{2H}^2 $

This condition insures the existence of infinite degenerate vacua with
$<H_1^0>\neq 0$ and $<H_2^0>\neq 0$.
\item[2.-] $m_{1H}^2+m_{2H}^2\geq 2|m_{12}|^2 $

This condition is needed to insure vacuum stability.
\end{itemize}
Once these conditions are imposed, the next step is to choose one out
of the infinite degenerate vacua. The usual asymmetric vacuum is the
simplest one which is defined by the following configuration:

\begin{center}
    $<H_1^0>=v_1\;;\; <H_2^0>=v_2$\\
    $v_1$ and $v_2$ are real and possitive\\
    $m_{12}^2$ is real and possitive
\end{center}

Furthermore, $v_1$ and $v_2$ are not completely free. They must fulfil
the following additional constraint:
\begin{equation}
m_W^2=\frac{1}{2}g^2(v_1^2+v_2^2) \Rightarrow v_1^2+v_2^2=(246\;GeV)^2
\label{const}
\end{equation}
One can use $m_{1H}^2$, $m_{2H}^2$ and $m_{12}^2$ as input parameters to
characterize the breaking or some three alternative parameters. For
instance, $v_1$, $v_2$ and the mass of the CP-odd scalar boson
$m_{A^0}$. If we consider, in addition, the constraint of
eq.(\ref{const}) we are left with just two independent papameters. It is
customary to choose as input parameters: $m_{A^0}$ and $\tan \beta
\equiv \frac{v_2}{v_1}$.

After some algebra one finds out the Higgs masses in terms of these
parameters:
\begin{eqnarray}
m_{H^{\pm}}^2&=&m_W^2+m_{A^0}^2\nonumber \\
m_{H_0,h_0}^2&=&\frac{1}{2}\left [ m_{A^0}^2+m_Z^2\pm \sqrt {
(m_{A^0}^2+m_Z^2)^2-4m_Z^2m_{A^0}^2\cos ^2 2 \beta } \right ]
\label{masses}
\end{eqnarray}
as well as the Higgs spectrum:
\begin{eqnarray}
H^+&;&H^- \nonumber \\
H^0&=&(\sqrt{2}Re\;H_1^0-v_1)\cos \alpha+(\sqrt{2}Re\; H_2^0-v_2)\sin
\alpha \nonumber \\
h^0&=&-(\sqrt{2}Re\;H_1^0-v_1)\sin \alpha+(\sqrt{2}Re\; H_2^0-v_2)\cos
\alpha
\end{eqnarray}
where,
\begin{equation}
\cos 2\alpha\equiv -\cos 2\beta\left (\frac{m_{A^0}^2-m_Z^2}
{m_{H^0}^2-m_{h_0}^2} \right )
 \;;\;
\sin 2\alpha\equiv -\sin 2\beta\left (\frac{m_{H^0}^2+m_{h^0}^2}
{m_{H^0}^2-m_{h_0}^2} \right )
\end{equation}
The result in eq.(\ref{masses}) indicates that the following
inequalities hold in the MSSM at tree level,
\begin{equation}
m_{h^0} \leq m_Z \;;\; m_{H^0} \geq m_Z \;;\; m_{H^{\pm}} \geq m_W
\label{ineq}
\end{equation}
Interestingly, there is a neutral Higgs boson $h_0$ lighter than the $Z$
boson. This fact, when it was noticed, seemed to announce a possible
discovery of $h_0$ at LEP. However, it was realized later that, beyond
tree level, these inequalities	in eq.(\ref{ineq}) do not hold
anylonger.
In particular, $m_h^0$ gets large corrections from top and stop loops
and one finds out to one loop that $m_{h^0}>m_Z$. By scanning the whole
parameter space, recent studies indicate, however, that the values
obtained for
$m_{h^0}$, including the complete one-loop corrections, never exceed
certain value. In ref.\cite{espin} the following absolute upper limit is
found:
\[ m_{h^0} < 140\; GeV \]
On the other hand, since these scalar particles have not been seen at
present experiments one can extract experimental lower mass limits.
From the absence of any Higgs signal at the LEP experiment one finds
\cite{susyex},
\begin{equation}
m_{h^0}^{\rm exp}>52\; GeV\;;\;
m_{A^0}^{\rm exp}>54\; GeV\;;\;
m_{H^{\pm}}^{\rm exp}>44\; GeV\;\;\;(95\%C.L.)
\end{equation}

\section{Strongly Interacting $\sbs$}
The strongly interacting hipothesis in the SM refers to the possibility
that the scalar self-coupling $\lambda$ be large and a perturbative
approach in powers of this coupling is not anylonger valid.
Since in the SM at tree level there is a direct relation between
$\lambda$ and $M_H$ given by $\lambda=\frac{g^2M_H^2}{8M_W^2}$, a large
value of $\lambda$ implies a large value of $M_H$. Thus, for instance,
for a very heavy Higgs with $M_H \sim 1\;TeV$ one gets a non
perturbative coupling of $\lambda\sim 7$.

Given the SM potential of eq.(\ref{lsbs}), a large value of $\lambda$
implies that the interactions among the three
would-be-Goldstone bosons and the Higgs particle are strong. Since, by
virtue of the Equivalence Theorem of eq.(\ref{et}), there is a relation
between the Goldstone bosons  and the longitudinal
gauge bosons scattering amplitudes,
\begin{equation}
T(V_L^1V_L^2\rightarrow V_L^3V_L^4)=T(w^1w^2\rightarrow
w^3w^4)+O(\frac{M_V^2}{s})\;;\;\sqrt{s}>>M_V\;;\;V^i=W^{\pm},Z
\end{equation}
it in turn implies that, at high enough energies, the $W_L^{\pm}$ and
$Z_L$ gauge bosons become strongly interacting too \cite{sis1,sis2}. The
amplitudes
of longitudinal gauge bosons are expected to show the typical features
of a strong interaction as, for instance, the appearance of multiple
resonances with sizeable widths in the TeV energy region etc.
\cite{sis2}. In the
case of the SM, the first resonance would be the Higgs particle itself
with a large mass and a large width \cite{screen}.

A common feature to all models of
strongly interacting $\sbs$ is that the size of the cross-section  for
production of
longitudinal gauge bosons pairs at these high energies, $\sqrt{s}\sim
O(1\;TeV)$, is expected to be larger than in weakly interacting theories
as, for
instance, the SM with a light Higgs boson. Although, there are several
interesting possibilities to look for strongly interacting signals, the
most obvious and, therefore, the most studied one is precisely $V_LV_L$
production at the future collider LHC in the various possible channels
,$V=W^{\pm}, Z$ \cite{sis2}. In the large energy region it comes mainly
from
the so-called gauge boson fusion processes (see Fig.7). Several studies
indicate that an enhacement in $V_LV_L$ production over the expected
background could be observed at LHC in the mass invariant region of
$M_{VV}\sim O(1\;TeV)$ \cite{sis2,sisex}.

Some comments are in order. There are some intrinsic problems connected
to the above definition of strongly interacting $\sbs$.
 One is
that for a too heavy Higgs particle, the width is comparable with the
mass and an interpretation of $H$ as a particle or as a resonance makes
no sense. Another one is that for such a heavy boson, the elastic
unitarity condition is violated in $V_LV_L$ scattering,
what indicates the failure of perturbation theory. In practice, some
unitarization procedure must, therefore,  be implemented to cure this
bad behaviour. In a more ambitious program, a non-perturbative treatment
of this strongly interacting system should be performed. The lattice,
would be obviously one possibility for future estimates.

Finally it is worth mentioning that one can also postulate the
hipothesis of a strongly interacting $\sbs$ beyond the SM. There are
several proposals and all have as a common assumption that either the
Higgs particle does not exist or it is not a fundamental particle. Some
examples are: Technicolor Models \cite{tc,tcrev}, models where the Higgs
boson is a top-antitop condensate \cite{ttb}, the BESS model
\cite{bess} etc..

\section{Low Energy Theorems}
These theorems are a consequence of the additional global symmetry,
\begin{equation}
\lr \rightarrow \lpr
\label{chiral}
\end{equation}
which is present in the $\sbs$ of the SM and we wish to be present
as well in any alternative Higgs sector in view of the successfull
prediction of $\rho=1$ based on this symmetry pattern.

The Low Energy Theorems are universal since they rely just on symmetry
arguments and, therefore, they must hold in any possible scenario for
the  $\sbs$. They state the following:

{\it The Goldstone boson scattering amplitudes	that
are imposed by the symmetry of eq.}(\ref{chiral}) {\it are given
at low energies by the following simple expressions} \cite{let1}:
\begin{eqnarray}
T(w^+w^-\rightarrow w^+w^-)&=&-\frac{u}{v^2}\nonumber \\
T(w^+w^-\rightarrow zz)&=& \frac{s}{v^2} \nonumber \\
T(zz\rightarrow zz)&=& 0
\label{let}
\end{eqnarray}
where, $v=246\;GeV$, ($s,t,u$) are the Mandelstan variables and low
energies here refer to energies well below any possible emerging
resonance.

It is interesting to notice the similarities with the Low Energy
Theorems for $\pi \pi$ scattering which are associated with the Chiral
symmetry of QCD \cite{let2}:
\begin{eqnarray}
T(\pi^+\pi^-\rightarrow \pi^+\pi^-)&=&-\frac{u}{f_{\pi}^2}\nonumber \\
T(\pi^+\pi^-\rightarrow \pi^0\pi^0)&=& \frac{s}{f_{\pi}^2} \nonumber \\
T(\pi^0 \pi^0 \rightarrow \pi^0 \pi^0)&=& 0
\end{eqnarray}
where, $f_{\pi}=94\; MeV$ and low energies here means
$\sqrt{s}<<m_{\rho}$.

Finally, by using the Equivalence Theorem, the above expressions in
eq.(\ref{let}) are
translated into Low Energy Theorems for the scattering of longitudinal
gauge bosons:
\begin{eqnarray}
T(W_L^+W_L^-\rightarrow W_L^+W_L^-)&=&-\frac{u}{v^2}\nonumber \\
T(W_L^+W_L^-\rightarrow Z_LZ_L)&=& \frac{s}{v^2} \nonumber \\
T(Z_LZ_L\rightarrow Z_LZ_L)&=& 0
\end{eqnarray}
They are universal as well and hold for any particular $\sbs$. The
energy must be in the range of applicability of both the Equivalence
Theorem and the Low Energy Theorems. It means an energy larger than the
$W^{\pm}$ and $Z$ masses but lower that the first possible resonance. In
the particular case of the SM, the above expressions should hold in the
energy range $M_{W,Z}<<\sqrt{s}<<M_H$. This is indeed what results from
the exact tree level expressions of eq.(\ref{long}) when this energy
limit is considered.

\section{Effective Lagrangian Approach to Electroweak Theory}
The electroweak interactions can be described at low energies by means
of an effective Lagrangian which is $\gs$ gauge invariant and is written
in terms of just the light fields \cite{effl}. In particular, the
effective
Lagrangian which does not contain explicitely the Higgs field in its
formulation  has been named Electroweak Chiral Lagrangian and has
some
interesting applications to electroweak phenomenology
\cite{sisex,echl1,echl2}. In this
approach the Higgs particle is assumed either very heavy, say $M_H \sim
O(1\;TeV)$, or unexistent.

In the bosonic sector, the EChL is a non-linear theory which is built in
terms of a field $U$ that parametrizes the three would-be-Godstone
bosons, its covariant derivative $D_{\mu}U$ and the $\gs$ gauge boson
fields \cite{applong}:
\begin{eqnarray}
U & \equiv & \displaystyle{\exp\left( {i \;
\frac{\vec{\tau}\cdot\vec{w}}{v}}\right)},\;\;\;
v  = 246 \;{\rm GeV}, \;\;\; \vec{w} = (w^1,w^2,w^3)
\nonumber\\
D_\mu U & \equiv & \partial_\mu
U - g \wh_\mu U + g' U \bh_\mu
\end{eqnarray}
where, $\wh_\mu$ and $\bh_\mu$ are defined in eq.(\ref{fields}).

It is a non-linear theory since the would-be-Goldstone bosons transforms
non-linearly under $\gs$ and it is a consequence of the lack of the
Higgs field in this theory that could complete together with the $w's$ a
linear multiplet. The $U$ field, however, transforms linearly under
$\gs$:
\begin{equation}
U(x) \rightarrow g_L U(x) g_Y^+ \;;\; g_L \in SU(2)_L\;;\;g_Y \in
U(1)_Y
\end{equation}
The EChL has the following generic form:
\begin{equation}
\ecl = \nll + \sum_{i=0}^{13} \cl_{i}. \label{ECL}
\end{equation}
where,
\begin{equation}
\nll  =  \frac{v^2}{4}\; Tr\left[ D_\mu U^\dagger D^\mu U \right]
+ \cl_{\rm YM}
\label{NLL}
\end{equation}
is the Lagrangian of the well known gauged non-linear sigma model and
$\cl_{\rm YM}$ is the Yang-Mills Lagrangian containing the kinetic, the
gauge fixing and the Faddeev Popov terms.

The $\cl_{i}$'s in eq.(\ref{ECL}) are the so-called chiral effective
operators. They are the complete set of operators with dimension up to
four (notice that the field $U$ is dimensionless) that can be built up
in terms of the light bosonic fields, $U$, $W^{\pm}_{\mu}$, $Z_{\mu}$
and $\gamma_{\mu}$, and that are $\gs$ and CP invariant. The building
blocks to implement gauge invariance are, therefore, the covariant
derivative $D_{\mu}U$ and the field strength tensors $\wtd$ and $\btd$
of eq.(\ref{tensors}).

For completeness, we include here the list of operators
\cite{applong}
\footnote{ The relation with Longhitano's notation in ref.\cite{applong}
is the following: $
a_0=\frac{g^2}{g'^2}\beta_1\;;\;a_1=\frac{g}{g'}\alpha_1\;;\;
a_2=\frac{g}{g'}\alpha_2\;;\;a_3=-\alpha_3\;;\;a_i=\alpha_i\;,
i=4,5,6,7\;;\;a_8=-\alpha_8\;;\;a_9=-\alpha_9\;;\;
a_{10}=\alpha_{10}/2\;;\;a_{11}=\alpha_{11}\;;\;
a_{12}=\alpha_{12}/2\;;\;a_{13}=\alpha_{13}$.  }:
\begin{eqnarray}
\cl_{0} & = & a_0 g'^2 \frac{v^2}{4} \left[ Tr\left(
T V_\mu \right) \right]^2 \nonumber\\[2mm]
\cl_{1} & = & a_1 \frac{i g g'}{2} B_{\mu\nu}
Tr\left( T \wtu \right) \nonumber\\[2mm]
\cl_{2} & = & a_2 \frac{i g'}{2} B_{\mu\nu}
Tr\left( T [V^\mu,V^\nu ] \right) \nonumber\\[2mm]
\cl_{3} & = & a_3  g Tr\left( \wtd [V^\mu,V^\nu ]\right)
\nonumber\\[2mm]
\cl_{4} & = & a_4  \left[ Tr\left( V_\mu V_\nu \right)
\right]^2 \nonumber\\[2mm]
\cl_{5} & = & a_5  \left[ Tr\left( V_\mu V^\mu \right)
\right]^2 \nonumber\\[2mm]
\cl_{6} & = & a_6 Tr\left( V_\mu V_\nu \right) Tr\left( T V^\mu
\right) Tr\left( T V^\nu \right)\nonumber\\[2mm]
\cl_{7} & = & a_7 Tr\left( V_\mu V^\mu \right) \left[
Tr\left( T V^\nu \right) \right]^2\nonumber\\[2mm]
\cl_{8} & = & a_8  \frac{g^2}{4} \left[ Tr\left( T \wtd \right)
\right]^2 \nonumber\\[2mm]
\cl_{9} & = & a_9  \frac{g}{2} Tr\left( T \wtd \right)
Tr\left( T [V^\mu,V^\nu ] \right) \nonumber\\[2mm]
\cl_{10} & = & a_{10} \left[ Tr\left( T V_\mu \right)
Tr\left( T V_\nu \right) \right]^2 \nonumber\\[2mm]
\cl_{11} & = & a_{11} Tr\left( ( D_\mu V^\mu )^2 \right)
\nonumber\\[2mm]
\cl_{12} & = & a_{12} Tr\left( T D_\mu D_\nu V^\nu \right)
Tr \left( T V^\mu \right)\nonumber\\[2mm]
\cl_{13} & = & a_{13} \frac{1}{2} \left[ Tr \left( T D_\mu V_\nu
\right) \right]^2 \label{Li}
\end{eqnarray}
where,
\begin{equation}
T  \equiv  U \tau^3 U^\dagger, \hspace{2cm}
 V_\mu\equiv(D_\mu U) U^\dagger .
\end{equation}
By reading from $\cl_{\rm NL}$ the quadratic terms in the gauge fields
one finds out the gauge boson masses:
\begin{equation}
M_W^2=\frac{g^2v^2}{4}\;;\;
M_Z^2=\frac{(g'^2+g^2)v^2}{4}\;;\;M_{\gamma}^2=0
\end{equation}
Therefore, the EChL describes as well a $\gs$ gauge theory with
spontaneous symmetry breaking to $U(1)_{\rm em}$. Furthermore, one can
check that the scalar sector of EChL has the additional global
symmetry $\lr$ and it is spontaneously broken down to the
custodial symmetry group $\lpr$. This is precisely the origen of
including the name 'chiral' in the EChL.

This approach to Electroweak Theory is inspired in the well
known Chiral Lagrangian approach to QCD at low energies and the Chiral
Perturbation Theory \cite{chpt}. In particular, the predictions in this
theory for
the would-be-Goldstone boson scattering amplitudes (in the approximation
of neglecting the gauge interactions versus the scalar
self-interactions) are similar to the pion scattering amplitudes of
Chiral Perturbation Theory \cite{echl1}. Furthermore, by virtue of the
Equivalence Theorem \cite{pel} it implies predictions for the scattering
amplitudes of the longitudinal gauge bosons.

Finally, the coefficients $a_i$ in front of the effective operators are
called the chiral parameters, and are very important since they encode
the information on the particular underlying physics which is the
responsible of generating this effective Lagrangian to low energies.
$\cl_{\rm NL}$ and the effective operators are universal whereas the
values of the chiral parameters do depend on the underlying assumed
fundamental theory. Therefore it is crucial to meassure in experiment
the $a_i$ values in order to be able to discriminate among the different
possible scenarios. On the other hand, it is also important to compute
these coefficients from the various posssible theories. Several works
have been done along these two lines. Some of the $a_i$'s can already be
bounded from present experiments as LEP (Holdon et al., Dobado et
al., Golden et al. in ref.\cite{echl2}; \cite{ailep}). In fact, one can
find
relations between some of these parameters and the S, T and U parameters
of ref.\cite{peskin} or, equivalently, the $\epsilon_i$ variables of
ref.\cite{alta} (see, for instance, ref.\cite{ester}) which have been
object of many studies in the last years.

The most chalenging
experiment for the study of the $\sbs$ with Effective Lagrangians will
be LHC. Recent studies indicate that the chiral parameters will be
measured (or bounded) mainly by analysing gauge boson pair production
processes in the high mass invariant region \cite{maite}. On the other
hand, there
are already available the computations of the chiral parameters in the
two most typical scenarios: The SM with a heavy Higgs particle
\cite{ester,ditt}, and Technicolor \cite{appwu}.

\section*{Acknowledgements}
I wish to thank my friends Teresa Rodrigo and Alberto Ruiz for
organizing this meeting which I found very interesting and
enjoyable.  I would like also to thank them for their 'infinite'
patience in waiting for these written lectures to be completed.
This
work has been partially supported
by the spanish Ministerio de Educaci\'on y Ciencia under project
CICYT AEN93-0673.


\end{document}